\documentclass[12pt]{article}

\usepackage{amsmath,amssymb,amsfonts,amsthm,hyperref}
\usepackage{graphicx,multirow,caption,bbm,latexsym,epsfig}
\usepackage{cite}

\newcommand{\bs}{\begin{subequations}}
\newcommand{\es}{\end{subequations}}
\newcommand{\be}{\begin{equation}}
\newcommand{\ee}{\end{equation}}
\newcommand{\ba}{\begin{eqnarray}}
\newcommand{\ea}{\end{eqnarray}}
\newcommand{\no}{\nonumber \\}

\newcommand{\vvv}{\mathcal{V}}

\allowdisplaybreaks

\begin{document}

\title{
\normalsize \hfill CFTP/21-007
\\[3mm] \LARGE
Fitting the $Z b \bar b$ vertex in the two-Higgs-doublet model
and in the three-Higgs-doublet model}

\author{
  Darius~Jur\v{c}iukonis$^{(1)}$\thanks{E-mail:
    \tt darius.jurciukonis@tfai.vu.lt}
  \
  and Lu\'\i s~Lavoura$^{(2)}$\thanks{E-mail:
    \tt balio@cftp.tecnico.ulisboa.pt}
  \\*[3mm]
  $^{(1)}\!$
  \small Vilnius University, Institute of Theoretical Physics and Astronomy, \\
  \small Saul\.etekio~ave.~3, Vilnius 10257, Lithuania
  \\*[2mm]
  $^{(2)}\!$
  \small Universidade de Lisboa, Instituto Superior T\'ecnico, CFTP, \\
  \small Av.~Rovisco~Pais~1, 1049-001~Lisboa, Portugal
}

\maketitle

\begin{abstract}
  We investigate the new contributions to the parameters $g_L$ and $g_R$
  of the $Z b \bar b$ vertex in a multi-Higgs-doublet model (MHDM).
  We emphasize that those contributions generally worsen
  the fit of those parameters to the experimental data.
  We propose a solution to this problem,
  wherein $g_R$ has the opposite sign
  from the one predicted by the Standard Model;
  this solution,
  though,
  necessitates light scalars and large Yukawa couplings in the MHDM.
\end{abstract}

\newpage

\section{Introduction}
\label{introduction}

In this
paper
we focus on the $Z b \bar b$ coupling
\be
\mathcal{L}_{Zbb} = \frac{g}{c_w}\, Z_\mu\, \bar b \gamma^\mu
\left( g_L P_L + g_R P_R \right) b,
\ee
where $c_w$ is the cosine of the weak mixing angle
and $P_L$ and $P_R$ are the projection operators of chirality.
At tree level,
\be
g_L^\mathrm{tree} = \frac{s_w^2}{3} - \frac{1}{2},
\quad
g_R^\mathrm{tree} = \frac{s_w^2}{3},
\ee
where $s_w$ is the sine of the weak mixing angle.
With $s_w^2 = 0.22337$~\cite{RPP},
one obtains $g_L^\mathrm{tree} = -0.42554$
and $g_R^\mathrm{tree} = 0.07446$.
The Standard Model (SM) prediction is~\cite{field}
\be
\label{SM}
g_L^\mathrm{SM} = -0.420875,
\quad
g_R^\mathrm{SM} = 0.077362.
\ee
In the presence of New Physics,
we write
\be
\label{delta}
g_L = g_L^\mathrm{SM} + \delta g_L,
\quad
g_R = g_R^\mathrm{SM} + \delta g_R.
\ee

Experimentally,
we get at $g_L$ and $g_R$ by measuring two quantities called $A_b$ and $R_b$;
their precise experimental definitions may be found
%%%%% in refs.~\cite{field,haber}.
in refs.~\cite{field,haber,freitas} and in appendix~\ref{appendixA}.
One has
\be
\label{v7fr895}
A_b = \frac{2 r_b \sqrt{1 - 4 \mu_b}}{1 - 4 \mu_b + \left( 1 + 2 \mu_b \right)
  r_b^2},
\ee
where $r_b = \left( g_L + g_R \right) \!
\left/ \left( g_L - g_R \right) \right.$
and $\mu_b = \left. \left[ m_b \left( m_Z^2 \right) \right]^2 \right/ m_Z^2$.
We use the numerical values
$m_b \left( m_Z^2 \right) = 3$\,GeV and $m_Z = 91.1876$\,GeV~\cite{RPP}.
Equation~\eqref{v7fr895} may be inverted to yield
\be
\label{vvuf9tr0}
\frac{g_L}{g_R} := \varrho =
\frac{\sqrt{1 - 4 \mu_b}
  \left[ 1 \pm \sqrt{1 - \left( 1 + 2 \mu_b \right) A_b^2} \right]
  + \left( 1 + 2 \mu_b \right) A_b}{\sqrt{1 - 4 \mu_b}
  \left[ 1 \pm \sqrt{1 - \left( 1 + 2 \mu_b \right) A_b^2} \right]
  - \left( 1 + 2 \mu_b \right) A_b}.
\ee
Notice the existence of two solutions for $\varrho$.
The other measured quantity is
\be
\label{rvufido}
R_b = \frac{s_b\, c^\mathrm{QCD}\, c^\mathrm{QED}}{s_b\,
  c^\mathrm{QCD}\, c^\mathrm{QED} + s_c + s_u + s_s + s_d},
\ee
where $c^\mathrm{QCD} = 0.9953$ and $c^\mathrm{QED} = 0.99975$
are QCD and QED corrections,
respectively,
\bs
\ba
\label{djfigf0}
s_b &=& \left( 1 - 6 \mu_b \right) \left( g_L - g_R \right)^2
+ \left( g_L + g_R \right)^2
\\ &=& g_R^2 \left[ \left( 2 - 6 \mu_b \right) \left( 1 + \varrho^2 \right)
  + 12 \mu_b \varrho \right], \label{djfigf0b}
\ea
\es
and $s_c + s_u + s_s + s_d = 1.3184$.
The solution to equations~\eqref{rvufido} and~\eqref{djfigf0b} is
\be
\label{vnfjoy}
g_R^2 = \frac{s_c + s_u + s_s + s_d}{c^\mathrm{QCD} c^\mathrm{QED}
\left[ \left( 2 - 6 \mu_b \right) \left( 1 + \varrho^2 \right)
  + 12 \mu_b \varrho \right]}\ \frac{R_b}{1 - R_b}.
\ee
Notice the two possible signs of $g_R$ in equation~\eqref{vnfjoy}.

%%%%% I exchanged the footnote by a reference.
An overall fit of many electroweak observables gives~\cite{freitas}
\bs
\label{dez}
\ba
R_b^\mathrm{fit} &=& 0.21629 \pm 0.00066, \label{Rbfit}
\\
A_b^\mathrm{fit} &=& 0.923 \pm 0.020. \label{Abfit}
\ea
\es
On the other hand,
$A_b$ has been directly measured at LEP1
and at SLAC in two different ways, see appendix~\ref{appendixA}.
The averaged result of those measurements is
\be
A_b^\mathrm{average} = 0.901 \pm 0.013. \label{Abtrue}
\ee
While the $A_b$ value of equation~\eqref{Abfit}
deviates from the Standard-Model $A_b$ value $0.9347$ by just 0.6$\sigma$,
the $A_b$ value of equation~\eqref{Abtrue}
displays a much larger disagreement of 2.6$\sigma$.

In this work we consider both the set of values~\eqref{dez},
which we denote through the superscript
``fit,''
and the set formed by the values~\eqref{Rbfit} and~\eqref{Abtrue},
which we denote through the superscript
``average.''
Plugging the central values of those two sets
into equations~\eqref{vvuf9tr0} and~\eqref{vnfjoy},
we obtain solutions~1,
2,
3,
and~4 for $g_L$ and $g_R$ in table~\ref{table_solutions}.
We also display in that table the corresponding values of
$\delta g_L = g_L + 0.420875$ and $\delta g_R = g_R - 0.077362$.
\begin{table}[ht]
  \begin{center}
    \begin{tabular}{|c|rr|rr|}
     \hline
     \multicolumn{1}{|c|}{solution} &
     \multicolumn{1}{c}{$g_L$} &
     \multicolumn{1}{c}{$g_R$} &
     \multicolumn{1}{|c}{$\delta g_L$} &
     \multicolumn{1}{c|}{$\delta g_R$} \\ \hline \hline
     1$^\mathrm{fit}$ & $-0.420206$ & $0.084172$ \hspace*{0.1mm} &
     \hspace*{0.1mm} $0.000669$ & $0.006810$ \\
     2$^\mathrm{fit}$ & $-0.419934$ & $-0.082806$ \hspace*{0.1mm} &
     \hspace*{0.1mm} $0.000941$ & $-0.160168$ \\
     3$^\mathrm{fit}$ & $0.420206$ & $-0.084172$ \hspace*{0.1mm} &
     \hspace*{0.1mm} $0.841081$ & $-0.161534$ \\
     4$^\mathrm{fit}$ & $0.419934$ & $0.082806$ \hspace*{0.1mm} &
     \hspace*{0.1mm} $0.840809$ & $0.005444$ \\
     \hline
     1$^\mathrm{average}$ & $-0.417814$ & $0.095496$ \hspace*{0.1mm} &
     \hspace*{0.1mm} $0.003061$ & $0.018134$ \\
     2$^\mathrm{average}$ & $-0.417504$ & $-0.094139$ \hspace*{0.1mm} &
     \hspace*{0.1mm} $0.003371$ & $-0.171501$ \\
     3$^\mathrm{average}$ & $0.417814$ & $-0.095496$ \hspace*{0.1mm} &
     \hspace*{0.1mm} $0.838688$ & $-0.172858$  \\
     4$^\mathrm{average}$ & $0.417504$ & $0.094139$ \hspace*{0.1mm} &
     \hspace*{0.1mm} $0.838379$ & $0.016777$ \\
     \hline
    \end{tabular}
  \end{center}
  \vspace*{-5mm}
  \caption{The results of equations~\eqref{vvuf9tr0} and~\eqref{vnfjoy}
    for $g_L$ and $g_R$ and the corresponding values of $\delta g_L$
    and $\delta g_R$ extracted from equations~\eqref{SM} and~\eqref{delta}.
    The superscript ``fit'' corresponds to the input values~\eqref{dez},
    while the superscript ``average'' corresponds
    to the input values~\eqref{Rbfit} and~\eqref{Abtrue}.
    \label{table_solutions}}
\end{table}
We see that solutions~3 and~4 have a much too large $\delta g_L$;
we outright discard those solutions.\footnote{Solutions~3 and~4 are good
  when one only measures $R_b$ and $A_b$ at the $Z^0$ peak;
  when one gets away from that peak,
  the diagram with an intermediate photon becomes significant
  and one easily finds that solutions~3 and~4
  are not really experimentally valid~\cite{tait}.
  So,
  there are both theoretical and experimental reasons
  for discarding them.}
Solution~1 seems to be preferred over solution~2
because it has much smaller
$\left| \delta g_R \right|$.\footnote{A recent preprint~\cite{yan}
  claims that there are already a couple LHC points
  that favour solution~1 over solution~2
  and that in the future the two solutions could be decisively discriminated
  through the high-luminosity-LHC data.
  On the other hand,
  the older ref.~\cite{tait} claims that the PETRA (35\,GeV) data
  actually favour solution~2 over solution~1.}
Still,
in this work we shall also consider solution~2.

In this paper we seek to reproduce solutions~1 and~2 by invoking New Physics,
specifically either the two-Higgs-doublet model (2HDM)
or the three-Higgs-doublet model (3HDM).
%%%%% I HAVE INSERTED HERE SOME REFERENCES.
The 2HDM is one of the simplest possible extensions of the SM. 
One of the many motivations for the 2HDM is supersymmetry:
the Minimal Supersymmetric Standard Model has two Higgs doublets.
Also,
the 2HDM may generate a Baryon Asymmetry of the Universe sufficiently large,
due to the flexibility of its scalar mass spectrum.
We recommend the review~\cite{Branco:2011iw} on the 2HDM in general, 
and refs.~\cite{Pich:2009sp,Pilaftsis:2016erj} on the aligned 2HDM.
In recent years the 3HDM has received increased attention,
see {\it e.g.}\ refs.\cite{Logan:2020mdz,Ivanov:2012fp,Keus:2013hya}.
The aligned 3HDM is discussed in refs.~\cite{Pilaftsis:2016erj,Das:2019yad}.

The plan of this work is as follows.
In section~\ref{nHDM} we present the general formulas
of $\delta g_L$ and $\delta g_R$ in
the
$n$-Higgs-doublet model.
In section~\ref{section:2HDM} we consider the particular case
of an aligned 2HDM and we specify the constraints on the scalar masses
that we have used in that case.
We do the same job for an aligned 3HDM in section~\ref{sec:3HDM}.
We then present numerical results in section~\ref{sec_numerics},
followed by our conclusions in section~\ref{sec:conclusions}.
Appendix~\ref{appendixA} deals on the definition of $R_b$ and $A_b$
and on the experimental data for them.
Appendix~\ref{appendixB}
works out the derivation
of the neutral-scalar contributions to $\delta g_L$ and $\delta g_R$.

\section{The $Z b \bar b$ vertex in the aligned $n$HDM}
\label{nHDM}

\subsection{Mixing formalism}

In a general $n$-Higgs-doublet model ($n$HDM) and utilizing,
without loss of generality,
the `charged Higgs basis'~\cite{bento},
the scalar doublets $\Phi_1, \ldots \Phi_n$ are written
\be
\label{doublets}
\Phi_1 = \left( \begin{array}{c}
  S_1^+ \\ \left( v + H + i S_1^0 \right) \left/ \sqrt{2} \right.
\end{array} \right),
\quad
\Phi_k = \left( \begin{array}{c}
  S_k^+ \\ \left( R_k + i I_k \right) \left/ \sqrt{2} \right.
\end{array} \right)
\quad
(k = 2, \ldots, n),
\ee
where $S_1^+$ is a charged Goldstone boson,
$S_1^0$ is the neutral Goldstone boson,
$v \approx 246$\,GeV is the (real and positive) vacuum expectation value (VEV),
and $S_2^+, \ldots, S_n^+$ are physical charged scalars
with masses $m_{C2}, \ldots, m_{Cn}$,
respectively.
%The matrix $U$ defined in equation~(15) of ref.~\cite{ogreid}
%is from the start equal to the $n \times n$ unit matrix.
Without loss of generality,
we order the doublets $\Phi_k$ through $m_{C2} \le m_{C3} \le \cdots \le m_{Cn}$.
We are free to rephase each of the $\Phi_k$,
thereby mixing $R_k$ and $I_k$ through a $2 \times 2$ orthogonal matrix.

The real fields $H$,
$R_k$,
and $I_k$ ($k = 2, \ldots, n$) are not eigenstates of mass,
rather
\be
\label{V}
\left( \begin{array}{c}
  H + i S_1^0 \\ R_2 + i I_2 \\ \vdots \\ R_n + i I_n
\end{array} \right)
= \vvv \left( \begin{array}{c} S_1^0 \\ S_2^0 \\ \vdots \\ S_{2n}^0
\end{array} \right),
\ee
where $\vvv$ is an $n \times 2n$ matrix
with $\left( 1,\, 1 \right)$ matrix element $\vvv_{11} = i$.
The physical neutral-scalar fields $S_2^0, \ldots, S_{2n}^0$ are real and
have masses $m_2, \ldots, m_{2n}$,
respectively.
An important property of $\vvv$ is that
\be
\label{tildeV}
\left( \begin{array}{c} \mathcal{R} \\ \mathcal{I} \end{array} \right) :=
\left( \begin{array}{c} \mathrm{Re}\, \vvv \\ \mathrm{Im}\, \vvv
\end{array} \right)
\ \mathrm{is\ a}\ 2n \times 2n\ \mathrm{real\ orthogonal\ matrix.}
\ee

For the sake of simplicity,
we assume alignment.
This means that $H \equiv S_2^0$ is a physical neutral scalar
that does \emph{not} mix with the $R_k$ and $I_k$.
Hence,
$\vvv_{12} = 1$ and $\vvv_{1j} = 0,\ \forall j = 3, \ldots, 2n$;
also,
$\vvv_{k1} = \vvv_{k2} = 0,\ \forall k = 2, \ldots, n$. 
The scalar $H$
is assumed to be the particle with mass $m_2 \approx 125$\,GeV
that has been observed at the LHC.
In this
paper,
alignment is just a simplifying assumption that we do not pretend to justify
through any symmetry imposed on the $n$HDM.
We order the $S_j^0$ through $m_3 \le m_4 \le \ldots \le m_{2n}$.
Notice that,
in principle,
one or more of these masses may be lower than $m_2$.

We define the real antisymmetric matrix
\be
\label{A}
\mathcal{A} := \mathrm{Im} \left( \vvv^\dagger \vvv \right)
= \mathcal{R}^T \mathcal{I} - \mathcal{I}^T \mathcal{R}
= \left( \begin{array}{ccccccc}
  0 & -1 & 0 & 0 & 0 & \ldots & 0 \\
  1 & 0 & 0 & 0 & 0 & \ldots & 0 \\
  0 & 0 & 0 & \mathcal{A}_{34} & \mathcal{A}_{35} & \ldots &
  \mathcal{A}_{3,2n} \\
  0 & 0 & - \mathcal{A}_{34} & 0 & \mathcal{A}_{45} & \ldots &
  \mathcal{A}_{4,2n} \\
  \vdots & \vdots & \vdots & \vdots & \vdots & \vdots & \vdots
\end{array} \right).
\ee

To compute the one-loop corrections to the $Z b \bar b$ vertex in the $n$HDM,
we make the simplifying assumption that only the top and bottom quarks exist
and the $\left( t,\, b \right)$
Cabibbo--Kobayashi--Maskawa matrix element is $1$.
The relevant part of the Yukawa Lagrangian is~\cite{fontes}
\be
\label{ef}
\mathcal{L}_\textrm{Yukawa} = -
\left( \begin{array}{cc} \overline{t_L} & \overline{b_L} \end{array} \right)
\sum_{k=2}^n \left[ \frac{f_k}{\sqrt{2}}
\left( \begin{array}{c} \sqrt{2}\, S_k^+ \\*[2mm]
R_k + i I_k \end{array} \right) b_R
+ \frac{e_k}{\sqrt{2}}
\left( \begin{array}{c} R_k - i I_k \\*[1mm] - \sqrt{2}\, S_k^-
\end{array} \right) t_R \right] + \textrm{H.c.},
\ee
where the $e_k$ and $f_k$ are Yukawa coupling constants.

\subsection{Passarino--Veltman functions}

%We use dimensional regularization,
%\textit{i.e.}\ the Feynman integrals are computed
%in a space--time of dimension $d = 4 - \epsilon$ with $\epsilon \to 0$.
The Passarino--Veltman function $B_1 \left( r^2, m_0^2, m_1^2 \right)$
is defined through
\be
%\mu^\epsilon \!
\int \! \frac{\mathrm{d}^4 k}{\left( 2 \pi \right)^4}\
\frac{1}{k^2 - m_0^2}\ \frac{1}{\left( k + r \right)^2 - m_1^2}\ k^\lambda
= \frac{i}{16 \pi^2}\ r^\lambda\,
B_1 \left( r^2, m_0^2, m_1^2 \right).
\label{b1}
\ee
The Passarino--Veltman function
$C_0 \left[ r_1^2, \left( r_1 - r_2 \right)^2, r_2^2,
  m_0^2, m_1^2, m_2^2 \right]$ is defined through
\ba
%\mu^\epsilon \!
\int \! \frac{\mathrm{d}^4 k}{\left( 2 \pi \right)^4}\
\frac{1}{k^2 - m_0^2}\
\frac{1}{\left( k + r_1 \right)^2 - m_1^2}\
\frac{1}{\left( k + r_2 \right)^2 - m_2^2}
&=&
\frac{i}{16 \pi^2}\ C_0 \left[ r_1^2, \left( r_1 - r_2 \right)^2, r_2^2, m_0^2,
  \right. \no & & \left.
  m_1^2, m_2^2 \right].
\label{c0}
\ea
The Passarino--Veltman functions $C_{00}$,
$C_{11}$,
$C_{22}$,
and $C_{12}$,
which depend on $r_1^2$,
$\left( r_1 - r_2 \right)^2$,
$r_2^2$,
$m_0^2$,
$m_1^2$,
and $m_2^2$ are defined through
\ba
%\mu^\epsilon \!
\int \! \frac{\mathrm{d}^4 k}{\left( 2 \pi \right)^4}\
\frac{1}{k^2 - m_0^2}\
\frac{1}{\left( k + r_1 \right)^2 - m_1^2}\
\frac{1}{\left( k + r_2 \right)^2 - m_2^2}\ k^\lambda k^\nu
&=&
\frac{i}{16 \pi^2} \left[ g^{\lambda \nu} C_{00}
+ r_1^\lambda r_1^\nu C_{11}
\right. \no & &
+ r_2^\lambda r_2^\nu C_{22}
+ \left( r_1^\lambda r_2^\nu
\right. \no & & \left. \left.
+ r_2^\lambda r_1^\nu \right) C_{12} \right]
\left[ r_1^2, \left( r_1 - r_2 \right)^2,
  \right. \no & & \left.
  r_2^2, m_0^2, m_1^2, m_2^2 \right].
\label{c00}
\ea
The functions $B_1 \left( r^2, m_0^2, m_1^2 \right)$
and $C_{00}\left[ r_1^2, \left( r_1 - r_2 \right)^2, r_2^2,
  m_0^2, m_1^2, m_2^2 \right]$ are divergent,
yet the functions $f_{L,R} \left( m^2 \right)$
and $h_{L,R} \left( m_j^2,\, m_{j^\prime}^2 \right)$
that are defined below in equations~\eqref{fLR} and~\eqref{hLR},
respectively,
are finite.

\subsection{The charged-scalar contribution}

\begin{figure}[ht]
  \begin{center}
    \includegraphics[width=0.8\textwidth]{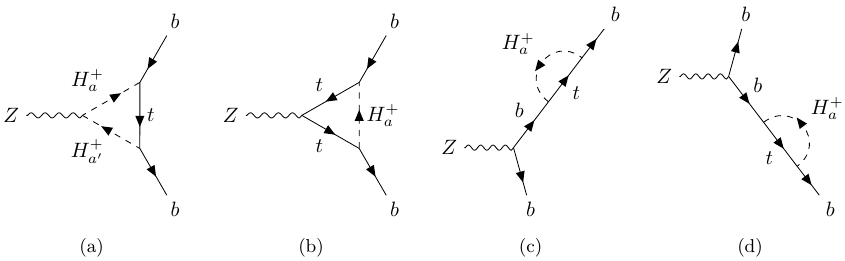}
  \end{center}
  \caption{Feynman diagrams that produce equations~\eqref{charged}
  and~\eqref{fLR}.}
\label{chargeddiagrams}
\end{figure}

\begin{figure}[ht]
  \begin{center}
    \includegraphics[width=0.6\textwidth]{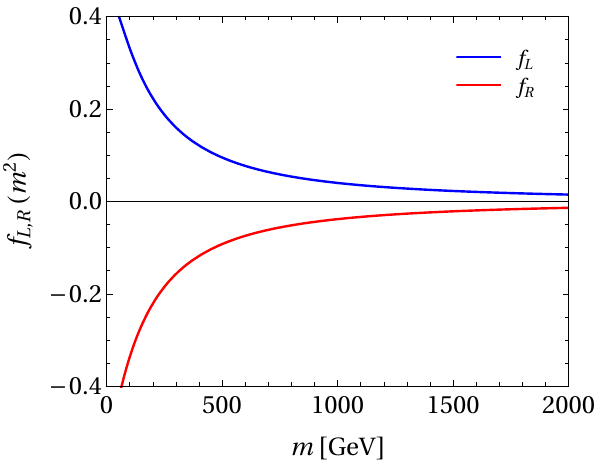}
  \end{center}
  \caption{The functions $f_L \left( m^2 \right)$ and $f_R \left( m^2 \right)$.}
\label{fig:fLR}
\end{figure}

In the $n$HDM at the one-loop level,
both $\delta g_L$ and $\delta g_R$ are the sum of a contribution,
which we denote through a superscript $c$,
from diagrams having charged scalars and top quarks in the internal lines
of the loop,
and another contribution,
which we denote through a superscript $n$,
from diagrams with neutral scalars and bottom quarks in the internal lines:
\be
\delta g_L = \delta g_L^c + \delta g_L^n,
\quad
\delta g_R = \delta g_R^c + \delta g_R^n.
\ee
The charged-scalar contribution has been computed long time ago~\cite{haber}.
%%%%% NEW TEXT
It corresponds to the computation of the diagrams
in figure~\ref{chargeddiagrams}.
It is
\be
\label{charged}
\delta g_L^c = \frac{1}{16 \pi^2} \sum_{k=2}^n \left| e_k \right|^2
f_L \left( m_{Ck}^2 \right),
\quad
\delta g_R^c = \frac{1}{16 \pi^2} \sum_{k=2}^n \left| f_k \right|^2
f_R \left( m_{Ck}^2 \right),
\ee
where the functions $f_L$ and $f_R$ are defined through
\bs
\label{fLR}
\ba
f_L \left( m^2 \right) &=&
\left( 2 s_w^2 - 1 \right)
C_{00} \left( 0,\, m_Z^2,\, 0,\, m_t^2,\, m^2,\, m^2 \right)
\no & &
+ \left( \frac{2 s_w^2}{3} - \frac{1}{2} \right)
m_t^2\, C_0 \left( 0,\, m_Z^2,\, 0,\, m^2,\, m_t^2,\, m_t^2 \right)
\no & &
- \frac{2 s_w^2}{3} \left[
  2\, C_{00} \left( 0,\, m_Z^2,\, 0,\, m^2,\, m_t^2,\, m_t^2 \right)
  - \frac{1}{2}
  \right. \no & & \left.
  - m_Z^2\, C_{12} \left( 0,\, m_Z^2,\, 0,\, m^2,\, m_t^2,\, m_t^2 \right)
  \vphantom{\frac{1}{2}} \right]
\no & &
+ \left( \frac{s_w^2}{3} - \frac{1}{2} \right)
B_1 \left( 0,\, m_t^2, m^2 \right),
\\
f_R \left( m^2 \right) &=&
\left( 2 s_w^2 - 1 \right)
C_{00} \left( 0,\, m_Z^2,\, 0,\, m_t^2,\, m^2,\, m^2 \right)
\no & &
+ \frac{2 s_w^2}{3}\,
m_t^2\, C_0 \left( 0,\, m_Z^2,\, 0,\, m^2,\, m_t^2,\, m_t^2 \right)
\no & &
+ \left( \frac{1}{2} - \frac{2 s_w^2}{3} \right) \left[
  2\, C_{00} \left( 0,\, m_Z^2,\, 0,\, m^2,\, m_t^2,\, m_t^2 \right)
  - \frac{1}{2}
  \right. \no & & \left.
  - m_Z^2\, C_{12} \left( 0,\, m_Z^2,\, 0,\, m^2,\, m_t^2,\, m_t^2 \right)
  \vphantom{\frac{1}{2}} \right]
\no & &
+ \frac{s_w^2}{3}\,
B_1 \left( 0,\, m_t^2, m^2 \right).
\ea
\es
In equations~\eqref{fLR} $m_t$ is the top-quark mass
and $m_Z$ is the mass of the gauge boson $Z^0$.
In the approximation $m_Z = 0$,
the functions $f_L$ and $f_R$ do not depend of $s_w$\footnote{When $m_Z = 0$
  the $Z^0$ is indistinguishable from the photon
  and therefore the weak mixing angle is arbitrary and unphysical.}
and are symmetric of each other:
\be
\label{vufifo}
f_R \left( m^2 \right)
\approx - f_L \left( m^2 \right) \approx \frac{1}{2}\,
\frac{x}{1 - x} \left( 1 + \frac{\ln{x}}{1 - x} \right),
\ee
where $x = m_t^2 / m^2$.
Remarkably,
the approximations~\eqref{vufifo} hold very well
even when one computes $f_L$ and $f_R$ with $m_Z = 91.1876$\,GeV.
The functions $f_L$ and $f_R$ are depicted
in figure~\ref{fig:fLR}.\footnote{We have performed the numerical computation
  of Passarino--Veltman functions by
  using the Fortran library Collier~\cite{Denner:2016kdg}
  through interface CollierLink~\cite{Patel:2015tea}.}
One sees that $f_L \left( m^2 \right) > 0$,
$f_R \left( m^2 \right) < 0$,
and $f_R \left( m^2 \right) \approx - f_L \left( m^2 \right)$
for all values of $m^2$.
Moreover,
the absolute values of both functions decrease
with increasing $m^2$.
Therefore,
$\delta g_L^c > 0$,
$\delta g_R^c < 0$,
and both $\delta g_L^c$ and $- \delta g_R^c$
are monotonically decreasing functions of the charged-scalar masses.

\subsection{The neutral-scalar contribution}

\begin{figure}[ht]
  \begin{center}
    \includegraphics[width=0.8\textwidth]{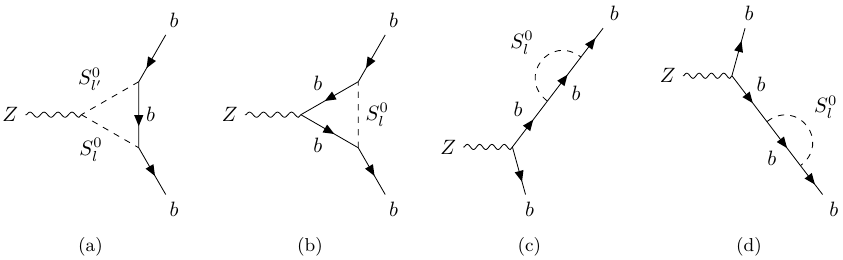}
  \end{center}
  \caption{Feynman diagrams that produce equations~\eqref{neutral}
  and~\eqref{hLR}.}
\label{neutraldiagrams}
\end{figure}

\begin{figure}[ht]
    \begin{center}
      \includegraphics[width=0.8\textwidth]{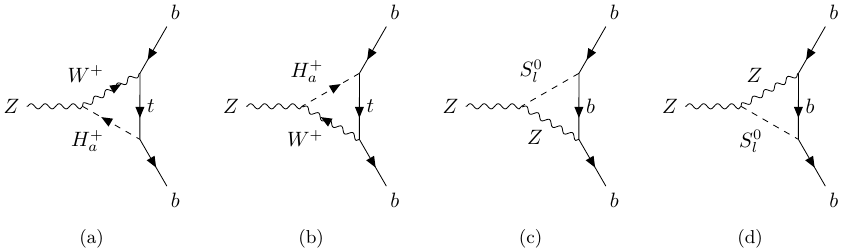}
    \end{center}
    \caption{Feynman diagrams that do not contribute to $\delta g_L$ and
      $\delta g_R$ in the specific case of this paper.}
    \label{nondiagrams}
  \end{figure}

\begin{figure}[ht]
  \begin{center}
    \includegraphics[width=1.0\textwidth]{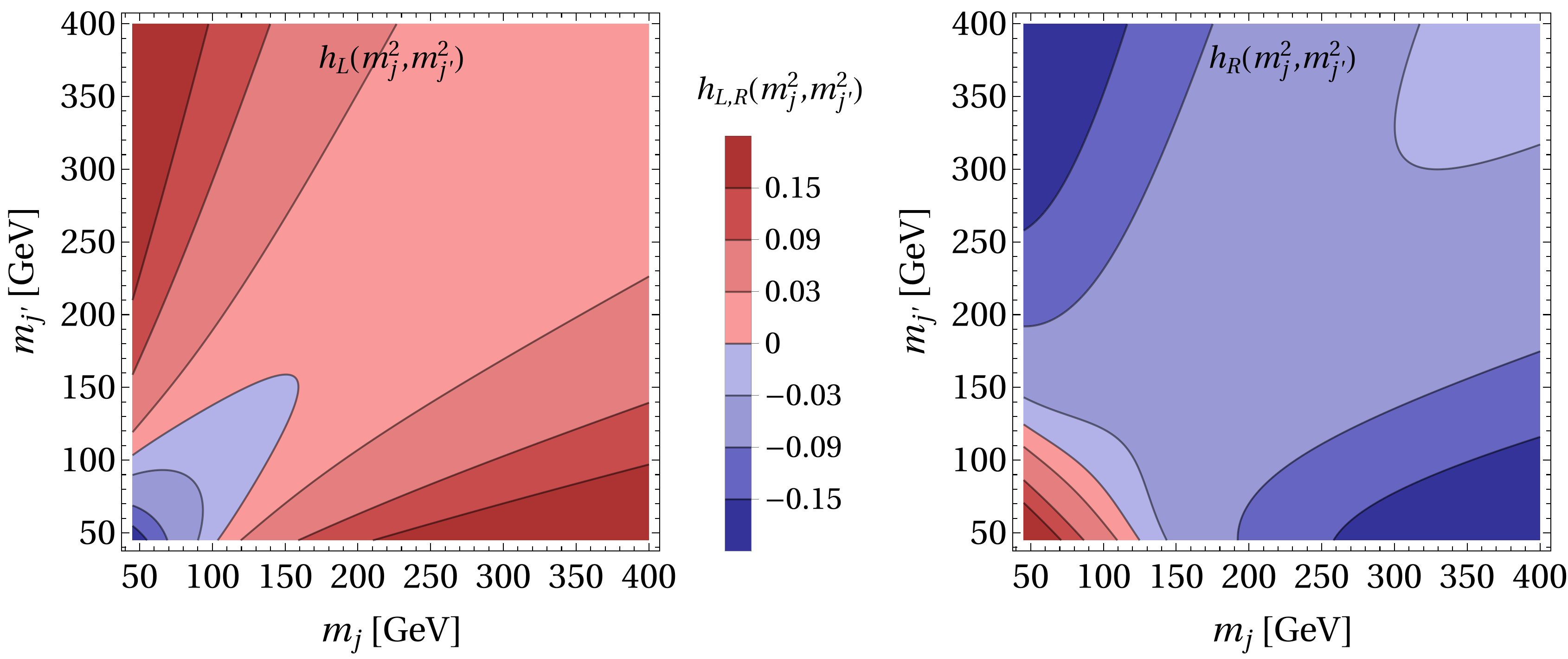}
  \end{center}
  \caption{The functions $h_L \left( m_j^2,\, m_{j^\prime}^2 \right)$
  and $h_R \left( m_j^2,\, m_{j^\prime}^2 \right)$.}
\label{figh}
\end{figure}

The neutral-scalar contribution to $\delta g_L$ and $\delta g_R$
has been recently emphasized in ref.~\cite{fontes},
following the original computation in ref.~\cite{haber};
%%%%% NEW TEXT
it corresponds to the computation of the diagrams
in figure~\ref{neutraldiagrams}
and
it is recapitulated
%%%%% NEW FOOTNOTE
in appendix~\ref{appendixB}.\footnote{The diagrams in figure~\ref{nondiagrams}
  do not contribute to $\delta g_L$ and $\delta g_R$ in our case.
  This is so because diagrams~(a) and~(b) only exist,
  if there are only scalar doublets,
  when $H_a^+$   is the charged Goldstone boson,
  and because diagrams~(c) and~(d) are proportional to the bottom-quark mass.}
Assuming alignment and discarding the Standard-Model contributions
that involve $S_1^0$ and $S_2^0$,
one has
\bs
\label{neutral}
\ba
\delta g_L^n &=& \frac{1}{16 \pi^2} \sum_{j=3}^{2n-1} \sum_{j^\prime=j+1}^{2n}
\mathcal{A}_{j j^\prime}\
\mathrm{Im} \left[ \left( \vvv^\dagger \mathcal{F}^\ast \right)_j
  \left( \vvv^T \mathcal{F} \right)_{j^\prime} \right]
h_L \left( m_j^2,\, m_{j^\prime}^2 \right),
\\
\delta g_R^n &=& \frac{1}{16 \pi^2} \sum_{j=3}^{2n-1} \sum_{j^\prime=j+1}^{2n}
\mathcal{A}_{j j^\prime}\
\mathrm{Im} \left[ \left( \vvv^\dagger \mathcal{F}^\ast \right)_j
  \left( \vvv^T \mathcal{F} \right)_{j^\prime} \right]
h_R \left( m_j^2,\, m_{j^\prime}^2 \right),
\ea
\es
where $\mathcal{F}$ is an $n \times 1$ vector
with
$k^\mathrm{th}$ component
$\mathcal{F}_k = f_k$ for $k = 2, \ldots, n$,
and
\bs
\label{hLR}
\ba
h_L \left( m_j^2,\, m_{j^\prime}^2 \right) &=&
- C_{00} \left( 0,\, m_Z^2,\, 0,\, 0,\, m_j^2,\, m_{j^\prime}^2 \right)
\no & &
+ \frac{s_w^2}{6} \left[
  2\, C_{00} \left( 0,\, m_Z^2,\, 0,\, m_j^2,\, 0,\, 0 \right)
  + 2\, C_{00} \left( 0,\, m_Z^2,\, 0,\, m_{j^\prime}^2,\, 0,\, 0 \right)
  \right. \no & & \left.
  - 1
  - m_Z^2\, C_{12} \left( 0,\, m_Z^2,\, 0,\, m_j^2,\, 0,\, 0 \right)
  - m_Z^2\, C_{12} \left( 0,\, m_Z^2,\, 0,\, m_{j^\prime}^2,\, 0,\, 0 \right)
  \right]
\no & &
+ \left( \frac{s_w^2}{6} - \frac{1}{4} \right) \left[
  B_1 \left( 0,\, 0,\, m_j^2 \right)
  + B_1 \left( 0,\, 0,\, m_{j^\prime}^2 \right) \right],
\\
h_R \left( m_j^2,\, m_{j^\prime}^2 \right) &=&
C_{00} \left( 0,\, m_Z^2,\, 0,\, 0,\, m_j^2,\, m_{j^\prime}^2 \right)
\no & &
+ \left( \frac{s_w^2}{6} - \frac{1}{4} \right) \left[
  2\, C_{00} \left( 0,\, m_Z^2,\, 0,\, m_j^2,\, 0,\, 0 \right)
  + 2\, C_{00} \left( 0,\, m_Z^2,\, 0,\, m_{j^\prime}^2,\, 0,\, 0 \right)
  \right. \no & & \left.
  - 1
  - m_Z^2\, C_{12} \left( 0,\, m_Z^2,\, 0,\, m_j^2,\, 0,\, 0 \right)
  - m_Z^2\, C_{12} \left( 0,\, m_Z^2,\, 0,\, m_{j^\prime}^2,\, 0,\, 0 \right)
  \right]
\no & &
+ \frac{s_w^2}{6} \left[
  B_1 \left( 0,\, 0,\, m_j^2 \right)
  + B_1 \left( 0,\, 0,\, m_{j^\prime}^2 \right) \right].
\ea
\es
The functions $h_L$ and $h_R$ are independent of $s_w$ when $m_Z = 0$;
however,
that approximation is not a good one for those functions.
We depict their real parts in figure~\ref{figh}.\footnote{The functions $h_L$
and $h_R$ are complex. However, their imaginary parts are irrelevant
for the computation of $g_L$ and $g_R$, since they do not interfere
with the tree-level contributions to those parameters~\cite{fontes},
which are real. Therefore, in this paper
whenever we talk about $h_L$ and $h_R$ we really mean just the real
parts of those two functions.}
One sees that,
when both $m_j$ and $m_{j^\prime}$ are larger than the Fermi scale,
$h_L \left( m_j^2,\, m_{j^\prime}^2 \right) > 0$
and $h_R \left( m_j^2,\, m_{j^\prime}^2 \right) < 0$.
However,
if both $m_j \lesssim 100$\,GeV and $m_{j^\prime} \lesssim 100$\,GeV,
then both $h_L \left( m_j^2,\, m_{j^\prime}^2 \right)$
and $h_R \left( m_j^2,\, m_{j^\prime}^2 \right)$ invert their usual signs.
Moreover,
$\left| h_L \left( m_j^2,\, m_{j^\prime}^2 \right) \right|$
and $\left| h_R \left( m_j^2,\, m_{j^\prime}^2 \right) \right|$
become rather large
either when $\left| m_j - m_{j^\prime} \right| \gtrsim  200$\,GeV
and one of the masses $\lesssim 50$\,GeV,
or when both $m_j$ and $m_{j^\prime} \lesssim 50$\,GeV.

\section{The aligned 2HDM} \label{section:2HDM}

In a two-Higgs-doublet model with alignment~\cite{fontes},
the doublet $\Phi_2$ may be rephased
so that $R_2 \equiv S_3^0$ and $I_2 \equiv S_4^0$
are the new physical neutral scalars.
Then,
\be
\vvv = \left( \begin{array}{cccc} i & 1 & 0 & 0 \\ 0 & 0 & 1 & i
\end{array} \right),
\ee
hence $\mathcal{A}_{34} = 1$ and $\left( \vvv^\dagger \mathcal{F}^\ast \right)_3
\left( \vvv^T \mathcal{F} \right)_4 = i \left| f_2 \right|^2$.
There are five New-Physics parameters
on which $\delta g_L$ and $\delta g_R$ depend:
the neutral-scalar masses $m_3$ and $m_4$,
the charged-scalar mass $m_{C2}$,
and the Yukawa couplings $e_2$ and $f_2$.
One has~\cite{fontes}
\bs
\ba
\delta g_L &=& \frac{\left| e_2 \right|^2 f_L \left( m_{C2}^2 \right)
  + \left| f_2 \right|^2 h_L \left( m_3^2,\, m_4^2 \right)}{16 \pi^2},
\\
\delta g_R &=& \frac{\left| f_2 \right|^2 \left[
    f_R \left( m_{C2}^2 \right) + h_R \left( m_3^2,\, m_4^2 \right)
    \right]}{16 \pi^2}.
\ea
\es

We now consider the scalar potential of the 2HDM~\cite{review},
\ba
V &=&
\mu_1\, \Phi_1^\dagger \Phi_1 + \mu_2\, \Phi_2^\dagger \Phi_2
+ \left( \mu_3\, \Phi_1^\dagger \Phi_2 + \mathrm{H.c.} \right)
\no & &
+ \frac{\lambda_1}{2} \left( \Phi_1^\dagger \Phi_1 \right)^2
+ \frac{\lambda_2}{2} \left( \Phi_2^\dagger \Phi_2 \right)^2
+ \lambda_3\, \Phi_1^\dagger \Phi_1\, \Phi_2^\dagger \Phi_2
+ \lambda_4\, \Phi_1^\dagger \Phi_2\, \Phi_2^\dagger \Phi_1
\no & &
+ \left[ \frac{\lambda_5}{2} \left( \Phi_1^\dagger \Phi_2 \right)^2
+ \lambda_6\, \Phi_1^\dagger \Phi_1\, \Phi_1^\dagger \Phi_2
+ \lambda_7\, \Phi_2^\dagger \Phi_2\, \Phi_1^\dagger \Phi_2
+ \mathrm{H.c.} \right].
\label{Vpot}
\ea
In the Higgs basis,
$\mu_1 = - \left. \lambda_1 v^2 \right/ 2$
and $\mu_3 = - \left. \lambda_6 v^2 \right/ 2$.
Because of alignment,
$\lambda_6$ (and $\mu_3$) are zero and
\be
\lambda_1 = \frac{m_2^2}{v^2} =
\left( \frac{125\, \mathrm{GeV}}{246\, \mathrm{GeV}} \right)^2 \approx 0.258.
\ee
From the masses of the scalars we compute
\be
\lambda_4 = \frac{m_3^2 + m_4^2 - 2 m_{C2}^2}{v^2},
\quad
\Lambda_5 = \frac{m_4^2 - m_3^2}{v^2},
\label{lambda45}
\ee
where $\Lambda_5 := \left| \lambda_5 \right|$.

The masses $m_{C2}$,
$m_3$,
and $m_4$ are not completely free,
because they must comply with unitarity (UNI)
and bounded-from-below (BFB) requirements~\cite{review}.
For the sake of simplicity,
in our analysis we assume $\lambda_2 = \lambda_7 = 0$.
We enforce the UNI conditions
\be
\lambda_4^2 < 64 \pi^2 - 8 \pi \lambda_1,
\quad
\Lambda_5^2 < 64 \pi^2 - 8 \pi \lambda_1
\label{condUNI1}
\ee
on the quantities~\eqref{lambda45}.
Additionally,
there are
\begin{itemize}
\item BFB conditions~\cite{review}
  \be
  \lambda_3 > 0, \quad \lambda_3 + \lambda_4 - \Lambda_5 > 0;
  \label{UNI}
  \ee
  \item UNI conditions~\cite{review}
    \be
    \begin{array}{lcl}
      \left| \lambda_3 \right| + \left| \lambda_4 \right| < 8 \pi, & &
      \left| \lambda_3 \right| + \Lambda_5 < 8 \pi, \quad
      \\*[1mm]
      \left| \lambda_3 + 2 \lambda_4 \right| + 3 \Lambda_5 < 8 \pi, & &
      \left( 2 \lambda_3 + \lambda_4 \right)^2 < 64 \pi^2 - 24 \pi \lambda_1;
    \end{array}
    \label{BFB}
    \ee
  \item the condition to avoid the situation of `panic vacuum',
    namely~\cite{PV}
  \be
  \left[ \left( \frac{m_{C2}^2}{v^2} + \frac{\lambda_4}{2} \right)^2
    - \frac{\Lambda_5^2}{4} \right] \left( \frac{m_{C2}^2}{v^2}
  - \frac{\lambda_3}{2}
  \right) > 0.
  \label{PV}
  \ee
\end{itemize}
After computing $\lambda_4$ and $\Lambda_5$ through equations~\eqref{lambda45}
and after checking inequalities~\eqref{condUNI1},
we verify whether there is any value of $\lambda_3$
that satisfies the inequalities~\eqref{UNI}--\eqref{PV};
if there is,
then the inputed masses $m_{C2}$,
$m_3$,
and $m_4$ are valid;
else,
they are not.

%%%%% I HAVE REARRANGED THIS PARAGRAPH ON THE OBLIQUE PARAMETERS.
We also compute the contribution of the new scalars
to the oblique parameter
\be
T = \frac{1}{16 \pi s_w^2 m_W^2} \left[
  F \left( m_{C2}^2,\ m_3^2 \right) + F \left( m_{C2}^2,\ m_4^2 \right)
  - F \left( m_3^2,\ m_4^2 \right) \right],
\ee
where $m_W = 80.4$\,GeV is the mass of the gauge bosons $W^\pm$ and
\be
F \left( A,\, B \right) = \left\{ \begin{array}{lcl}
  \displaystyle{\frac{A + B}{2} - \frac{A B}{A - B}\, \ln{\frac{A}{B}}}
  & \Leftarrow & A \neq B,
  \\*[3mm]
  0 & \Leftarrow & A = B.
  \end{array} \right.
\ee
Additionally,
we apply constraints on the oblique parameter~\cite{oblique}
\be
\label{1}
S = \frac{4 s_w^2 c_w^2}{\alpha} \left[
  \left. \frac{\partial A_{ZZ} \left( q^2 \right)}{\partial q^2}
  \right|_{q^2 = m_Z^2}
  - \left. \frac{\partial A_{\gamma \gamma} \left( q^2 \right)}{\partial q^2}
  \right|_{q^2 = 0}
  + \frac{c_w^2 - s_w^2}{c_w s_w}
  \left. \frac{\partial A_{\gamma Z} \left( q^2 \right)}{\partial q^2}
  \right|_{q^2 = 0}
  \right].
\ee
This parameter has been computed in ref.~\cite{ogreid2} to be
\be
S = \frac{1}{24 \pi} \left[ \ln{\frac{m_3^2 m_4^2}{m_{C2}^4}}
+ \left( s_w^2 - c_w^2 \right)^2 f \left( m_{C2}^2,\, m_{C2}^2,\, m_Z^2 \right)
+ f \left( m_3^2,\, m_4^2,\, m_Z^2 \right) \right].
\ee
Here,
\bs
\label{f}
\ba
f \left( A,\, B,\, C \right) &=&
- \frac{10}{3} - 4\, \frac{A + B}{C} + 4\, \frac{\left( A - B \right)^2}{C^2}
+ \left[ 3\, \frac{A^2 - B^2}{C^2} - 2\, \frac{\left( A - B \right)^3}{C^3}
  \right] \ln{\frac{A}{B}}
\hspace*{7mm}
\\ & &
+ \left[ \frac{1}{C} + \frac{A + B}{C^2}
  - 2\, \frac{\left( A - B \right)^2}{C^3}
  \right] h \left( t,\, r \right),
\ea
\es
where $t \equiv A + B - C$,
$r \equiv A^2 + B^2 + C^2 - 2 \left( A B + A C + B C \right)$,
and
\be
\label{h}
h \left( t,\, r \right) =
\left\{ \begin{array}{lcl}
  {\displaystyle \sqrt{r}\,
    \ln{\left| \frac{t - \sqrt{r}}{t + \sqrt{r}} \right|}}
  &\Leftarrow& r > 0,
  \\
  0 & \Leftarrow & r = 0,
  \\
  {\displaystyle 2\, \sqrt{- r}\, \arctan{\frac{\sqrt{- r}}{t}}}
  &\Leftarrow& r < 0.
\end{array} \right.
\ee
We either enforce the phenomenological constraint~\cite{RPP}
\bs
\label{ST}
\ba
T &=& 0.03 \pm 0.12, \label{T} \\
S &=& -0.01 \pm 0.10 \label{S}
\ea
\es
or we allow for other New Physics beyond the 2HDM
and apply milder requirements by allowing values of $T$ and $S$
within its $3\sigma$ and $2\sigma$ bounds,
respectively.

\section{The aligned 3HDM} \label{sec:3HDM}

\subsection{Parameterization of the neutral-scalar mixing}

In the three-Higgs-doublet model with alignment,
\be
\label{vjf8f}
\left( \begin{array}{c} R_2 \\ R_3 \\ I_2 \\ I_3 \end{array} \right)
= \mathcal{T}
\left( \begin{array}{c} S_3^0 \\ S_4^0 \\ S_5^0 \\ S_6^0 \end{array} \right),
\ee
where $\mathcal{T}$ is a $4 \times 4$ real orthogonal matrix.
%Without lack of generality we set $\det{\mathcal{T}} = +1$.
We parameterize
\be
\label{Tparam}
\mathcal{T} =
\mathcal{O}_{13} \left( \theta_5 \right) \times
\mathcal{O}_{24} \left( \theta_6 \right) \times
\mathcal{O}_{12} \left( \theta_1 \right) \times
\mathcal{O}_{34} \left( \theta_2 \right) \times
\mathcal{O}_{14} \left( \theta_3 \right) \times
\mathcal{O}_{23} \left( \theta_4 \right),
\ee
where $\mathcal{O}_{pq} \left( \theta \right)$
represents a rotation through an angle $\theta$
in the $\left( p,\, q \right)$ plane.
%$\left[ \mathcal{O}_{pq} \left( \theta \right) \right]_{qp}
%= - \left[ \mathcal{O}_{pq} \left( \theta \right) \right]_{pq} = \sin{\theta}$
%and $\left[ \mathcal{O}_{pq} \left( \theta \right) \right]_{pp}
%= \left[ \mathcal{O}_{pq} \left( \theta \right) \right]_{qq} = \cos{\theta}$;
%in all its columns and rows but for the $p$'th and $q$'th ones,
%$\mathcal{O}_{pq} \left( \theta \right)$ is like the unit matrix.
Now,
$\mathcal{O}_{13} \left( \theta_5 \right)$ is a rotation that mixes
$R_2$ and $I_2$,
and
$\mathcal{O}_{24} \left( \theta_6 \right)$ is a rotation mixing $R_3$ and $I_3$,
\textit{viz.}\ they represent rephasings of the doublets $\Phi_2$ and $\Phi_3$,
respectively.
Since such rephasings are unphysical,
one may without loss of generality drop those two rotations
from the parameterization~\eqref{Tparam},
obtaining
\be
\label{bur9e8}
\mathcal{T} =
\left( \begin{array}{cccc}
  c_1 c_3 & - s_1 c_4 & s_1 s_4 & - c_1 s_3 \\ 
  s_1 c_3 & c_1 c_4 & - c_1 s_4 & - s_1 s_3 \\ 
  - s_2 s_3 & c_2 s_4 & c_2 c_4 & - s_2 c_3 \\ 
  c_2 s_3 & s_2 s_4 & s_2 c_4 & c_2 c_3
\end{array} \right),
\ee
where $c_p = \cos{\theta_p}$ and $s_p = \sin{\theta_p}$ for $p = 1, 2, 3, 4$.
Then,
\bs
\label{vk0r9}
\ba
\delta g_L &=& \frac{\left| e_2 \right|^2 f_L \left( m_{C2}^2 \right)
  + \left| e_3 \right|^2 f_L \left( m_{C3}^2 \right)}{16 \pi^2}
\no & &
  + \frac{1}{16 \pi^2} \sum_{j=3}^5 \sum_{j^\prime=j+1}^6 \mathcal{A}_{j j^\prime}\
  \mathrm{Im} \left[ \left( \mathcal{V}^\dagger \mathcal{F}^\ast \right)_j
  \left( \mathcal{V}^T \mathcal{F} \right)_{j^\prime} \right]
  h_L \left( m_j^2,\, m_{j^\prime}^2 \right),
\\
\delta g_R &=& \frac{\left| f_2 \right|^2 f_R \left( m_{C2}^2 \right)
  + \left| f_3 \right|^2 f_R \left( m_{C3}^2 \right)}{16 \pi^2}
\no & &
  + \frac{1}{16 \pi^2} \sum_{j=3}^5 \sum_{j^\prime=j+1}^6 \mathcal{A}_{j j^\prime}\
  \mathrm{Im} \left[ \left( \mathcal{V}^\dagger \mathcal{F}^\ast \right)_j
  \left( \mathcal{V}^T \mathcal{F} \right)_{j^\prime} \right]
  h_R \left( m_j^2,\, m_{j^\prime}^2 \right),
\ea
\es
with
\bs
\ba
\mathcal{A}_{34} = - \mathcal{A}_{56}
&=& \left( c_1 c_2 + s_1 s_2 \right) \left( c_3 s_4 - s_3 c_4 \right),
\\
\mathcal{A}_{35} = \mathcal{A}_{46}
&=& \left( c_1 c_2 + s_1 s_2 \right) \left( c_3 c_4 + s_3 s_4 \right),
\\
\mathcal{A}_{36} = - \mathcal{A}_{45}
&=& s_1 c_2 - c_1 s_2,
\ea
\es
and
\bs
\ba
\mathrm{Im} \left[ \left( \mathcal{V}^\dagger \mathcal{F}^\ast \right)_3
\left( \mathcal{V}^T \mathcal{F} \right)_4 \right] &=&
\left| f_2 \right|^2 \left( c_1 c_2 c_3 s_4 - s_1 s_2 s_3 c_4 \right)
+ \left| f_3 \right|^2 \left( s_1 s_2 c_3 s_4 - c_1 c_2 s_3 c_4 \right)
\no & &
+ \mathrm{Re} \left( f_2 f_3^\ast \right)
\left( c_1 s_2 + s_1 c_2 \right) \left( c_3 s_4 + s_3 c_4 \right)
\no & &
+ \mathrm{Im} \left( f_2 f_3^\ast \right)
\left( s_3 s_4 - c_3 c_4 \right),
\\
\mathrm{Im} \left[ \left( \mathcal{V}^\dagger \mathcal{F}^\ast \right)_3
\left( \mathcal{V}^T \mathcal{F} \right)_5 \right] &=&
\left| f_2 \right|^2 \left( c_1 c_2 c_3 c_4 + s_1 s_2 s_3 s_4 \right)
+ \left| f_3 \right|^2 \left( s_1 s_2 c_3 c_4 + c_1 c_2 s_3 s_4 \right)
\no & &
+ \mathrm{Re} \left( f_2 f_3^\ast \right)
\left( c_1 s_2 + s_1 c_2 \right) \left( c_3 c_4 - s_3 s_4 \right)
\no & &
+ \mathrm{Im} \left( f_2 f_3^\ast \right)
\left( c_3 s_4 + s_3 c_4 \right),
\\
\mathrm{Im} \left[ \left( \mathcal{V}^\dagger \mathcal{F}^\ast \right)_3
\left( \mathcal{V}^T \mathcal{F} \right)_6 \right] &=&
- c_1 s_2 \left| f_2 \right|^2 + s_1 c_2 \left| f_3 \right|^2
+ \mathrm{Re} \left( f_2 f_3^\ast \right)
\left( c_1 c_2 - s_1 s_2 \right),
\\
\mathrm{Im} \left[ \left( \mathcal{V}^\dagger \mathcal{F}^\ast \right)_4
\left( \mathcal{V}^T \mathcal{F} \right)_5 \right] &=&
- s_1 c_2 \left| f_2 \right|^2 + c_1 s_2 \left| f_3 \right|^2
+ \mathrm{Re} \left( f_2 f_3^\ast \right)
\left( c_1 c_2 - s_1 s_2 \right),
\\
\mathrm{Im} \left[ \left( \mathcal{V}^\dagger \mathcal{F}^\ast \right)_4
\left( \mathcal{V}^T \mathcal{F} \right)_6 \right] &=&
\left| f_2 \right|^2 \left( s_1 s_2 c_3 c_4 + c_1 c_2 s_3 s_4 \right)
+ \left| f_3 \right|^2 \left( c_1 c_2 c_3 c_4 + s_1 s_2 s_3 s_4 \right)
\no & &
- \mathrm{Re} \left( f_2 f_3^\ast \right)
\left( c_1 s_2 + s_1 c_2 \right) \left( c_3 c_4 - s_3 s_4 \right),
\no & &
- \mathrm{Im} \left( f_2 f_3^\ast \right)
\left( c_3 s_4 + s_3 c_4 \right),
\\
\mathrm{Im} \left[ \left( \mathcal{V}^\dagger \mathcal{F}^\ast \right)_5
\left( \mathcal{V}^T \mathcal{F} \right)_6 \right] &=&
\left| f_2 \right|^2 \left( c_1 c_2 s_3 c_4 - s_1 s_2 c_3 s_4 \right)
+ \left| f_3 \right|^2 \left( s_1 s_2 s_3 c_4 - c_1 c_2 c_3 s_4 \right)
\no & &
+ \mathrm{Re} \left( f_2 f_3^\ast \right)
\left( c_1 s_2 + s_1 c_2 \right) \left( c_3 s_4 + s_3 c_4 \right),
\no & &
+ \mathrm{Im} \left( f_2 f_3^\ast \right)
\left( s_3 s_4 - c_3 c_4 \right).
\ea
\es

The contribution of the new scalars to the oblique parameter $T$,
given in equation~(23) of ref.~\cite{ogreid},
is
\ba
T &=& \frac{1}{16 \pi s_w^2 m_W^2} \left\{
  \left( c_1^2 c_3^2 + s_2^2 s_3^2 \right) F \left( m_{C2}^2,\, m_3^2 \right)
  + \left( s_1^2 c_4^2 + c_2^2 s_4^2 \right) F \left( m_{C2}^2,\, m_4^2 \right)
  \right. \no & &
  \hspace*{18mm}
  + \left( s_1^2 s_4^2 + c_2^2 c_4^2 \right) F \left( m_{C2}^2,\, m_5^2 \right)
  + \left( c_1^2 s_3^2 + s_2^2 c_3^2 \right) F \left( m_{C2}^2,\, m_6^2 \right)
  \nonumber \\*[2mm] & &
  \hspace*{18mm}
  + \left( s_1^2 c_3^2 + c_2^2 s_3^2 \right) F \left( m_{C3}^2,\, m_3^2 \right)
  + \left( c_1^2 c_4^2 + s_2^2 s_4^2 \right) F \left( m_{C3}^2,\, m_4^2 \right)
  \nonumber \\*[2mm] & &
  \hspace*{18mm}
  + \left( c_1^2 s_4^2 + s_2^2 c_4^2 \right) F \left( m_{C3}^2,\, m_5^2 \right) 
  + \left( s_1^2 s_3^2 + c_2^2 c_3^2 \right) F \left( m_{C3}^2,\, m_6^2 \right)
  \nonumber \\*[1mm] & &
  \hspace*{18mm}
  - \left( \mathcal{A}_{34} \right)^2
  \left[ F \left( m_3^2,\, m_4^2 \right) + F \left( m_5^2,\, m_6^2 \right)
    \right]
  \nonumber \\*[1mm] & &
  \hspace*{18mm}
  - \left( \mathcal{A}_{35} \right)^2
  \left[ F \left( m_3^2,\, m_5^2 \right) + F \left( m_4^2,\, m_6^2 \right)
    \right]
  \nonumber \\*[1mm] & & \left.
  \hspace*{18mm}
  - \left( \mathcal{A}_{36} \right)^2
  \left[ F \left( m_3^2,\, m_6^2 \right) + F \left( m_4^2,\, m_5^2 \right)
    \right]
  \right\}.
  \label{8g}
\ea
For the oblique parameter $S$ one has
\bs
\ba
S &=& \frac{1}{24 \pi} \left\{
  \left( s_w^2 - c_w^2 \right)^2 \left[
    f \left( m_{C2}^2,\, m_{C2}^2,\, m_Z^2 \right)
    + f \left( m_{C3}^2,\, m_{C3}^2,\, m_Z^2 \right)
    \right]
  \right. \\ & &
  +   \ln{\frac{m_3^2 m_4^2 m_5^2 m_6^2}{m_{C2}^4 m_{C3}^4}}
  + \left( \mathcal{A}_{34} \right)^2 \left[
    f \left( m_3^2,\, m_4^2,\, m_Z^2 \right)
    + f \left( m_5^2,\, m_6^2,\, m_Z^2 \right) \right]
  \\ & &
  + \left( \mathcal{A}_{35} \right)^2 \left[
    f \left( m_3^2,\, m_5^2,\, m_Z^2 \right)
    + f \left( m_4^2,\, m_6^2,\, m_Z^2 \right) \right]
  \\ & & \left.
  + \left( \mathcal{A}_{45} \right)^2 \left[
    f \left( m_3^2,\, m_6^2,\, m_Z^2 \right)
    + f \left( m_4^2,\, m_5^2,\, m_Z^2 \right) \right]
  %  \vphantom{\ln{\frac{m_3^2 m_4^2 m_5^2 m_6^2}{m_{C2}^4 m_{C3}^4}}}
  \right\}.
\ea
\es

\subsection{The scalar potential}

\paragraph{The parameters}
The scalar potential of the 3HDM has lots of couplings
and it is impractical to work with it.
So we concentrate on a truncated version of the potential,
\textit{viz.}\ we discard from the quartic part of the potential
all the terms that either do not contain $\Phi_1$
or are linear in $\Phi_1$.\footnote{This is equivalent to
  discarding from the scalar potential of the 2HDM
  the terms with coefficients $\lambda_2$ and $\lambda_7$,
  like we did in the previous section.}
The remaining potential is
\ba
\label{pote}
V &=&
\mu_1\, \Phi_1^\dagger \Phi_1
+ \mu_2\, \Phi_2^\dagger \Phi_2
+ \mu_3\, \Phi_3^\dagger \Phi_3
+ \left(
\mu_4\, \Phi_1^\dagger \Phi_2
+ \mu_5\, \Phi_1^\dagger \Phi_3
+ \mu_6\, \Phi_2^\dagger \Phi_3
+ \mathrm{H.c.} \right)
\no & &
+ \frac{\lambda_1}{2} \left( \Phi_1^\dagger \Phi_1 \right)^2
+ \lambda_4\, \Phi_1^\dagger \Phi_1\, \Phi_2^\dagger \Phi_2
+ \lambda_5\, \Phi_1^\dagger \Phi_1\, \Phi_3^\dagger \Phi_3
+ \lambda_7\, \Phi_1^\dagger \Phi_2\, \Phi_2^\dagger \Phi_1
+ \lambda_8\, \Phi_1^\dagger \Phi_3\, \Phi_3^\dagger \Phi_1
\no & &
+ \left[
  \frac{\lambda_{10}}{2} \left( \Phi_1^\dagger \Phi_2 \right)^2
  + \frac{\lambda_{11}}{2} \left( \Phi_1^\dagger \Phi_3 \right)^2
+ \lambda_{13}\, \Phi_1^\dagger \Phi_1\, \Phi_1^\dagger \Phi_2
+ \lambda_{14}\, \Phi_1^\dagger \Phi_1\, \Phi_1^\dagger \Phi_3
\right.
\no & & \left.
+ \lambda_{19}\, \Phi_1^\dagger \Phi_1\, \Phi_2^\dagger \Phi_3
+ \lambda_{22}\, \Phi_1^\dagger \Phi_3\, \Phi_2^\dagger \Phi_1
+ \lambda_{25}\, \Phi_1^\dagger \Phi_2\, \Phi_1^\dagger \Phi_3
+ \mathrm{H.c.} \vphantom{\frac{\lambda_{10}}{2}} \right],
\ea
where $\mu_{1,2,3}$ and $\lambda_{1,4,5,7,8}$ are real
and the remaining parameters are in general complex.
In order that the VEV of $\Phi_1$ is $v \left/ \sqrt{2} \right.$
and the VEVs of $\Phi_2$ and $\Phi_3$ are zero,
one must have
\be
\label{Higgs1}
\mu_1 = - \frac{\lambda_1 v^2}{2}, \quad
\mu_4 = - \frac{\lambda_{13} v^2}{2}, \quad
\mu_5 = - \frac{\lambda_{14} v^2}{2}.
\ee
In order that the charged-scalar mass matrix is
$\mathrm{diag} \left( m_{C2}^2,\, m_{C3}^2 \right)$,
one must have
\be
\label{Higgs2}
\mu_2 = m_{C2}^2 - \frac{\lambda_4 v^2}{2}, \quad
\mu_3 = m_{C3}^2 - \frac{\lambda_5 v^2}{2}, \quad
\mu_6 = - \frac{\lambda_{19} v^2}{2}.
\ee
Equations~\eqref{Higgs1} and~\eqref{Higgs2} are the conditions
for the charged Higgs basis.
Next we write down the conditions for alignment,
\textit{i.e.}\ for $H \equiv S_2^0$ to have mass $m_2$
and not to have mass terms together with either $R_2$,
$R_3$,
$I_2$,
or $I_3$:
\be
\label{1314}
\lambda_1 = \frac{m_2^2}{v^2} \approx 0.258,
\quad
\lambda_{13} = \lambda_{14} = 0,
\ee
hence $\mu_4$ and $\mu_5$ are zero too.
The mass terms of $R_2$,
$R_3$,
$I_2$,
and $I_3$ are given by
\be
V = \cdots + \frac{1}{2} \left( \begin{array}{cccc} R_2, & R_3, & I_2, & I_3
\end{array} \right) N \left( \begin{array}{c} R_2 \\ R_3 \\ I_2 \\ I_3
\end{array} \right),
\ee
where $N$ is a $4 \times 4$ real symmetric matrix.
Using equations~\eqref{vjf8f} and~\eqref{bur9e8},
one finds that
\bs
\label{vudieoe}
\ba
N_{11} - m_{C2}^2 & &
\no
= \frac{v^2}{2} \left( \lambda_7 + \mathrm{Re}\, \lambda_{10} \right) &=&
m_3^2 c_1^2 c_3^2 + m_4^4 s_1^2 c_3^2 + m_5^2 s_2^2 s_3^2 + m_6^2 c_2^2 s_3^2
- m_{C2}^2,
\\
N_{33} - m_{C2}^2 & &
\no
= \frac{v^2}{2} \left( \lambda_7 - \mathrm{Re}\, \lambda_{10} \right) &=&
m_3^2 s_1^2 s_4^2 + m_4^4 c_1^2 s_4^2 + m_5^2 c_2^2 c_4^2 + m_6^2 s_2^2 c_4^2
- m_{C2}^2,
\\
N_{13} = - \frac{v^2}{2}\, \mathrm{Im}\, \lambda_{10} &=&
\left( m_3^2 - m_4^2 \right) c_1 s_1 c_3 s_4
+ \left( m_6^2 - m_5^2 \right) c_2 s_2 s_3 c_4,
\\
N_{22} - m_{C3}^2 & &
\no
= \frac{v^2}{2} \left( \lambda_8 + \mathrm{Re}\, \lambda_{11} \right) &=&
m_3^2 s_1^2 c_4^2 + m_4^4 c_1^2 c_4^2 + m_5^2 c_2^2 s_4^2 + m_6^2 s_2^2 s_4^2
- m_{C3}^2,
\\
N_{44} - m_{C3}^2 & &
\no
= \frac{v^2}{2} \left( \lambda_8 - \mathrm{Re}\, \lambda_{11} \right) &=&
m_3^2 c_1^2 s_3^2 + m_4^4 s_1^2 s_3^2 + m_5^2 s_2^2 c_3^2 + m_6^2 c_2^2 c_3^2
- m_{C3}^2,
\\
N_{24} = - \frac{v^2}{2}\, \mathrm{Im}\, \lambda_{11} &=&
\left( m_3^2 - m_4^2 \right) c_1 s_1 s_3 c_4
+ \left( m_6^2 - m_5^2 \right) c_2 s_2 c_3 s_4,
\\
N_{12} = \frac{v^2}{2}\,
\mathrm{Re} \left( \lambda_{22} + \lambda_{25} \right) &=&
\left( m_4^2 - m_3^2 \right) c_1 s_1 c_3 c_4
+ \left( m_6^2 - m_5^2 \right) c_2 s_2 s_3 s_4,
\\
N_{34} = \frac{v^2}{2}\,
\mathrm{Re} \left( \lambda_{22} - \lambda_{25} \right) &=&
\left( m_4^2 - m_3^2 \right) c_1 s_1 s_3 s_4
+ \left( m_6^2 - m_5^2 \right) c_2 s_2 c_3 c_4,
\\
N_{14} = - \frac{v^2}{2}\,
\mathrm{Im} \left( \lambda_{22} + \lambda_{25} \right) &=&
c_3 s_3 \left( - c_1^2 m_3^2 - s_1^2 m_4^2 + s_2^2 m_5^2 + c_2^2 m_6^2 \right),
\\
N_{23} = \frac{v^2}{2}\,
\mathrm{Im} \left( \lambda_{22} - \lambda_{25} \right) &=&
c_4 s_4 \left( - s_1^2 m_3^2 - c_1^2 m_4^2 + c_2^2 m_5^2 + s_2^2 m_6^2 \right).
\ea
\es
Equations~\eqref{vudieoe} allow one to compute $\lambda_7$,
$\lambda_8$,
$\lambda_{10}$,
$\lambda_{11}$,
$\lambda_{22}$,
and $\lambda_{25}$ by using as input the masses of the charged scalars
and the masses and mixings of the neutral scalars.
On the other hand,
$\lambda_4$,
$\lambda_5$,
and $\lambda_{19}$ constitute extra parameters that we input by hand---just
as we did with $\lambda_3$ in section~\ref{section:2HDM}.

\paragraph{UNI constraints} These constraints state that
the moduli of the eigenvalues of some matrices must be smaller than $8 \pi$.
The method for the derivation of those matrices in a general $n$HDM
was explained in ref.~\cite{bento}.
In our specific case,
the UNI constraints are
(using $\Lambda_i \equiv \left| \lambda_i \right|$
for $i = 10, 11, 19, 22, 25$),
\bs
\label{vcjfuido}
\ba
\left| \lambda_4 + \lambda_5 - \lambda_7 - \lambda_8 \right|
+ \sqrt{\left( \lambda_4 - \lambda_5 - \lambda_7 + \lambda_8 \right)^2
  + 4 \left| \lambda_{19} - \lambda_{22} \right|^2} &<& 16 \pi,
\\
\left| \lambda_4 + \lambda_5 + \lambda_7 + \lambda_8 \right|
+ \sqrt{\left( \lambda_4 - \lambda_5 + \lambda_7 - \lambda_8 \right)^2
  + 4 \left| \lambda_{19} + \lambda_{22} \right|^2} &<& 16 \pi,
\\
\lambda_1 + \sqrt{\lambda_1^2 +
  4 \left( \Lambda_{10}^2 + \Lambda_{11}^2 + 2 \Lambda_{25}^2 \right)} &<& 16 \pi,
\\
\lambda_1 + \sqrt{\lambda_1^2 + 4 \left( \lambda_7^2 + \lambda_8^2
  + 2 \Lambda_{22}^2
  \right)} &<& 16 \pi,
\\
3 \lambda_1 + \sqrt{9 \lambda_1^2 +
  4 \left[ \left( 2 \lambda_4 + \lambda_7 \right)^2
  + \left( 2 \lambda_5 + \lambda_8 \right)^2
  + 2 \left| 2 \lambda_{19} + \lambda_{22} \right|^2 \right]} &<& 16 \pi,
\ea
\es
and the moduli of the eigenvalues of
\be
\left( \begin{array}{cccc}
  \lambda_4 & \lambda_{10} & \lambda_{19}^\ast & \lambda_{25} \\
  \lambda_{10}^\ast & \lambda_4 & \lambda_{25}^\ast & \lambda_{19} \\
  \lambda_{19} & \lambda_{25} & \lambda_5 & \lambda_{11} \\
  \lambda_{25}^\ast & \lambda_{19}^\ast & \lambda_{11}^\ast & \lambda_5
\end{array} \right)
\quad \mbox{and} \quad
\left( \begin{array}{cccc}
  \lambda_4 + 2 \lambda_7 & 3 \lambda_{10} &
  \lambda_{19}^\ast + 2 \lambda_{22}^\ast & 3 \lambda_{25} \\
  3 \lambda_{10}^\ast & \lambda_4 + 2 \lambda_7 &
  3 \lambda_{25}^\ast & \lambda_{19} + 2 \lambda_{22} \\
  \lambda_{19} + 2 \lambda_{22} & 3 \lambda_{25} &
  \lambda_5 + 2 \lambda_8 & 3 \lambda_{11} \\
  3 \lambda_{25}^\ast & \lambda_{19}^\ast + 2 \lambda_{22}^\ast &
  3 \lambda_{11}^\ast & \lambda_5 + 2 \lambda_8
\end{array} \right)
\ee
must be smaller than $8 \pi$.
In the inequalities~\eqref{vcjfuido},
all the square roots are taken positive
and $\lambda_1 = m_2^2 / v^2$ is positive too.

\paragraph{Necessary conditions for boundedness-from-below (BFB)}
The quartic part of the potential,
call it $V_4$,
must be positive for all possible configurations of the scalar doublets,
else the potential will be unbounded from below.
In the configuration $\Phi_3 = 0$,
the 3HDM
becomes
a 2HDM and one may use the BFB conditions
for the 2HDM~\cite{review}:
\be
\lambda_4 > 0, \quad \lambda_4 + \lambda_7 - \Lambda_{10} > 0.
\ee
Similarly,
from the configuration $\Phi_2 = 0$,
\be
\lambda_5 > 0, \quad \lambda_5 + \lambda_8 - \Lambda_{11} > 0.
\ee
We also consider the configuration
\be
\Phi_1 = \left( \begin{array}{c} \varphi_1 \\ 0 \end{array} \right), \quad
\Phi_2 = \left( \begin{array}{c} 0 \\ \varphi_2 \end{array} \right), \quad
\Phi_3 = \left( \begin{array}{c} 0 \\ \varphi_3 \end{array} \right),
\ee
wherein $\Phi_1^\dagger \Phi_2 = \Phi_1^\dagger \Phi_3 = 0$
but $\Phi_2^\dagger \Phi_3 \neq 0$.
Then,
\be
V_4 \ge \frac{\lambda_1}{2}\, r_1^2 + r_1
\left( \lambda_4 r_2 + \lambda_5 r_3 - 2 \Lambda_{19} \sqrt{r_2 r_3} \right),
\ee
where $r_q := \Phi_q^\dagger \Phi_q$ for $q = 1, 2, 3$.
By forcing $\lambda_4 r_2 + \lambda_5 r_3
- 2 \Lambda_{19} \sqrt{r_2 r_3}$ to be positive
for every positive $r_2$ and $r_3$,
one obtains the necessary BFB condition
\be
\lambda_4 + \lambda_5 - \sqrt{\left( \lambda_4 - \lambda_5 \right)^2
  + 4\, \Lambda_{19}^2} > 0.
\ee

\paragraph{Sufficient BFB conditions}
We know that $\Phi_q^\dagger \Phi_q\, \Phi_{q^\prime}^\dagger \Phi_{q^\prime}
- \Phi_q^\dagger \Phi_{q^\prime}\, \Phi_{q^\prime}^\dagger \Phi_q \ge 0$
when $q \neq {q^\prime}$.
Therefore,
we may parameterize
\be
\Phi_1^\dagger \Phi_2 = \sqrt{r_1 r_2}\, k_{12} e^{i \phi_{12}}, \quad
\Phi_1^\dagger \Phi_3 = \sqrt{r_1 r_3}\, k_{13} e^{i \phi_{13}}, \quad
\Phi_2^\dagger \Phi_3 = \sqrt{r_2 r_3}\, k_{23} e^{i \phi_{23}},
\ee
where $k_{12}$,
$k_{13}$,
and $k_{23}$ are real numbers in the interval $\left[0,\, 1 \right]$.
Then,
\bs
\ba
V_4 &=& \frac{\lambda_1}{2}\, r_1^2 + \lambda_4 r_1 r_2 + \lambda_5 r_1 r_3
+ \lambda_7 r_1 r_2 k_{12}^2 + \lambda_8 r_1 r_3 k_{13}^2
+ \left\{
\frac{\lambda_{10}}{2}\, r_1 r_2 k_{12}^2 e^{2 i \phi_{12}}
\right. \no & &
+ \frac{\lambda_{11}}{2}\, r_1 r_3 k_{13}^2 e^{2 i \phi_{13}}
+ r_1 \sqrt{r_2 r_3} \left[
  \lambda_{19} k_{23} e^{i \phi_{23}}
  + \lambda_{22} k_{12} k_{13} e^{i \left( \phi_{13} - \phi_{12} \right)}
  \right. \no & & \left. \left.
  + \lambda_{25} k_{12} k_{13} e^{i \left( \phi_{13} + \phi_{12} \right)}
  \right] + \mathrm{c.c.}
\vphantom{\frac{\lambda_{10}}{2}} \right\}
\\ &\ge& \frac{\lambda_1}{2}\, r_1^2 + \lambda_4 r_1 r_2 + \lambda_5 r_1 r_3
+ \left( \lambda_7 - \Lambda_{10} \right) r_1 r_2 k_{12}^2
+ \left( \lambda_8 - \Lambda_{11} \right) r_1 r_3 k_{13}^2
\\ & &
- 2 r_1 \sqrt{r_2 r_3} \left( \Lambda_{19} k_{23} + \Lambda k_{12} k_{13}
\right)
\\ &\ge& \frac{\lambda_1}{2}\, r_1^2 + \lambda_4 r_1 r_2 + \lambda_5 r_1 r_3
+ \left( \lambda_7 - \Lambda_{10} \right) r_1 r_2 k_{12}^2
+ \left( \lambda_8 - \Lambda_{11} \right) r_1 r_3 k_{13}^2
\\ & &
- r_1 \left( r_2 + r_3 \right)
\left( \Lambda_{19} k_{23} + \Lambda k_{12} k_{13} \right)
\\ &\ge&
\frac{\lambda_1}{2}\, r_1^2
+ r_1 r_2 \left[ \lambda_4 - \Lambda_{19}
  + \left( \lambda_7 - \Lambda_{10} \right) k_{12}^2
  - \Lambda\, k_{12} k_{13} \right]
\\ & &
\hspace*{10.7mm} + r_1 r_3 \left[ \lambda_5 - \Lambda_{19}
  + \left( \lambda_8 - \Lambda_{11} \right) k_{13}^2
  - \Lambda\, k_{12} k_{13} \right]
\\ &\ge& \frac{\lambda_1}{2}\, r_1^2
+ r_1 r_2 \left[ \lambda_4 - \Lambda_{19}
  + \left( \lambda_7 - \Lambda_{10} \right) k_{12}^2 - \Lambda\, k_{12} \right]
\\ & &
\hspace*{10.7mm} + r_1 r_3 \left[ \lambda_5 - \Lambda_{19}
  + \left( \lambda_8 - \Lambda_{11} \right) k_{13}^2 - \Lambda\, k_{13} \right].
\ea
\es
where $\Lambda := \Lambda_{22} + \Lambda_{25}$.
Thus,
denoting $L_7$ and $L_8$ the minimum values of
$\left( \lambda_7 - \Lambda_{10} \right) k_{12}^2 - \Lambda\, k_{12}$
and $\left( \lambda_8 - \Lambda_{11} \right) k_{13}^2 - \Lambda\, k_{13}$,
respectively,
one has the sufficient BFB conditions~\cite{ivanovferreira}
\be
\lambda_4 - \Lambda_{19} + L_7 > 0
\quad \mbox{and} \quad
\lambda_5 - \Lambda_{19} + L_8 > 0.
\ee
It is easy to find that
\bs
\ba
L_7 &=& \left\{ \begin{array}{lcl}
  \lambda_7 - \Lambda_{10} - \Lambda
  & \Leftarrow & \displaystyle{\lambda_7 - \Lambda_{10} < \frac{\Lambda}{2},}
  \\*[2mm]
  \displaystyle{- \frac{\Lambda^2}{4 \left( \lambda_7 - \Lambda_{10} \right)}}
  & \Leftarrow & \displaystyle{\lambda_7 - \Lambda_{10}
    > \frac{\Lambda}{2};}
\end{array} \right.
\\
L_8 &=& \left\{ \begin{array}{lcl}
  \lambda_8 - \Lambda_{11} - \Lambda
  & \Leftarrow & \displaystyle{\lambda_8 - \Lambda_{11} < \frac{\Lambda}{2},}
  \\*[2mm]
  \displaystyle{- \frac{\Lambda^2}{4 \left( \lambda_8 - \Lambda_{11} \right)}}
  & \Leftarrow & \displaystyle{\lambda_8 - \Lambda_{11} > \frac{\Lambda}{2}.}
\end{array} \right.
\ea
\es

\section{Numerical results} \label{sec_numerics}

In this section we display various scatter plots obtained by using
the formulas in sections~\ref{section:2HDM} and~\ref{sec:3HDM}.
In all the plots,
we have restricted the Yukawa couplings $e_2$,
$e_3$,
$f_2$,
and $f_3$ to have moduli smaller than $4 \pi$.
The charged-scalar masses $m_{C2}$ and $m_{C3}$ were
assumed to be
between 150\,GeV and 2\,TeV.
The neutral-scalar masses $m_3, \ldots, m_6$  were
supposed
to be lower than 2\,TeV,
but they have sometimes been allowed to be as low as 50\,GeV.
In practice,
the upper bound on the scalar masses is mostly irrelevant,
since the contributions of the new scalars to $\delta g_L$ and $\delta g_R$
tend to zero when the scalars become very heavy.

%%%%% I HAVE INSERTED HERE SOME REMARKS ON THE SCALAR MASSES.
Constraints on the masses of the new scalars of the 2HDM and 3HDM
may be derived from collider experiments
on the production and subsequent decay of on-shell Higgs bosons. 
The sensitivity is limited by the kinematic reach of the experiments;
moreover,
the constraints usually depend on the assumed Yukawa couplings
of the scalars and the fermions,
which we do not want to specify in our work.

Constraints from the process $b \to s \gamma$ are most stringent.
For a 2HDM of type II,
a lower bound $m_{C2} > 480$\,GeV at 95\% CL
has been derived in ref.~\cite{Misiak:2015xwa}.
However,
in the case of the 3HDM,
it has been shown in refs.~\cite{Logan:2020mdz,Akeroyd:2016ssd} that,
due to the increased number of parameters,
$m_{C2}$ and $m_{C3}$ may actually be lighter than the mass of top quark
while complying with the constraints from $b \to s \gamma$.

In the 2HDM,
a bound $m_{C2} \gtrsim 150$\,GeV on the mass of the charged Higgs boson
has been derived from searches
at the LHC~\cite{Khachatryan:2015qxa,Arbey:2017gmh}. 
Recent global fits~\cite{Chowdhury:2017aav,Eberhardt:2020dat}
give bounds on the scalar masses
for various types of Yukawa couplings in the 2HDM. 
%In these studies there is only considered 2HDM scenarios
%with a  $\mathbb{Z}_2$  symmetric potential
%but is possible to assume that proposed bounds
%could be similar to the general 2HDM also. 
In ref.~\cite{Chowdhury:2017aav} it is claimed that
the mass of the heavy CP-even Higgs boson $m_{H} > \{450,\, 700\}$~\,GeV,
the mass of the CP-odd Higgs boson $m_{A} > \{500,\, 750\}$\,GeV,
and the mass of the charged Higgs boson $m_{C2} > \{460,\, 740\}$\,GeV;
the first values in the curled brackets
correspond to the ``lepton specific'' type of 2HDM
while the second values correspond to the type II and the ``flipped'' 2HDM.
In the fit~\cite{Eberhardt:2020dat} of the aligned 2HDM
one finds a lower bound of the new-scalar masses $m_{C2}$,
$m_3$,
and $m_4$ around 500\,GeV or around 750\,GeV,
depending on the fitted mass range. 

\begin{figure}[ht]
  \begin{center}
    \includegraphics[width=1.0\textwidth]{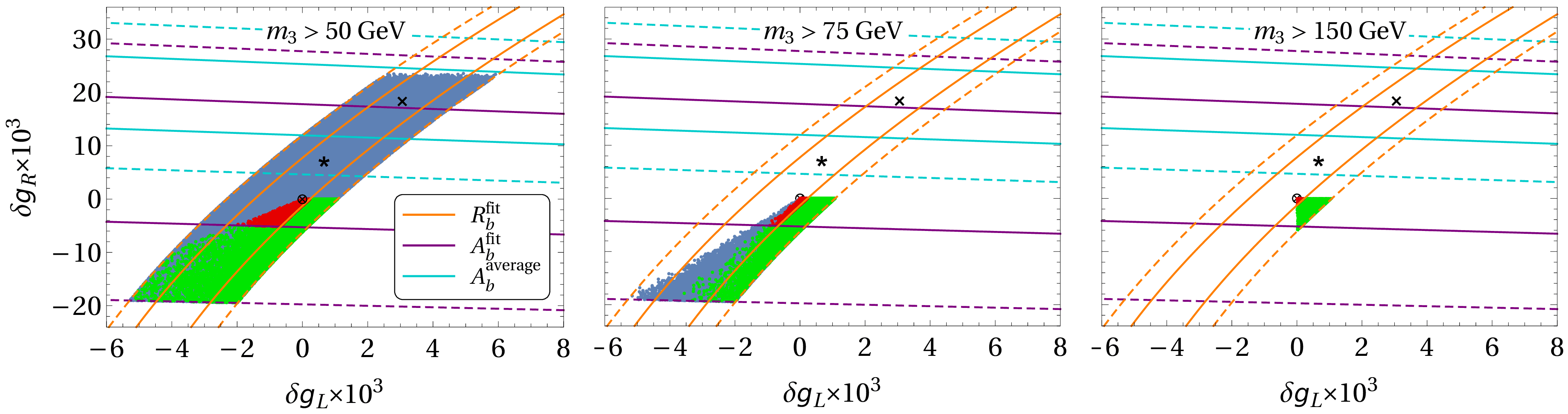}
  \end{center}
  \caption{Scatter plot of values of $\delta g_L$ and $\delta g_R$
    in the aligned 2HDM.
    A crossed circle marks the point $\delta g_L = \delta g_R = 0$.
    A star marks the best-fit point of solution~1$^\mathrm{fit}$,
    and a cross the best-fit point of solution~1$^\mathrm{average}$.
    The orange lines mark the $1\sigma$ (full lines)
    and $2\sigma$ (dashed lines) boundaries of the region determined
    by the experimental value~\eqref{Rbfit};
    similarly,
    the violet lines correspond to the value~\eqref{Abfit}
    and the light-blue lines to the value~\eqref{Abtrue}.
    Red points agree with the $1\sigma$ intervals~\eqref{dez};
    green points agree with the $2 \sigma$,
    but not with the $1\sigma$,
    intervals~\eqref{dez};
    and blue points agree either with the $1\sigma$ or the $2\sigma$
    intervals~\eqref{dez}.
    Both the red and the green points satisfy the $1\sigma$ limits
    in equation~\eqref{ST},
    while the blue points comply with laxer conditions where
    $S$ can reach $2\sigma$ bounds and $T$ can reach $3\sigma$ bounds
    in equation~\eqref{ST}.
    Note that some blue points are underneath either red or green points.
    Left panel: both new neutral scalars have masses above 50\,GeV;
    middle panel: both new neutral scalars have masses above 75\,GeV;
    right panel: both new neutral scalars have masses above 150\,GeV.
  }
\label{gLR2HDM}
\end{figure}

\begin{figure}[ht]
  \begin{center}
    \includegraphics[width=1.0\textwidth]{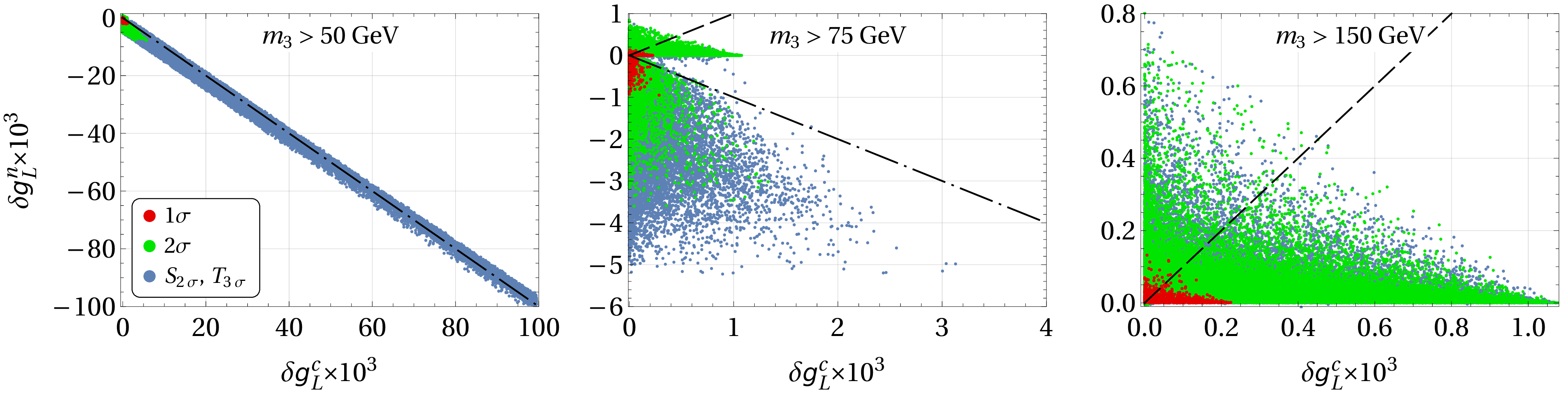}
  \end{center}
  \caption{Scatter plot of $\delta g_L^n$ \textit{versus}\/ $\delta g_L^c$
    in the aligned 2HDM.
    The displayed points and the colour code employed
    are the same as in figure~\ref{gLR2HDM}.
    The dashed straight lines mark the condition $\delta g_L^n = \delta g_L^c$
    and the dashed-dotted lines correspond to $\delta g_L^n = - \delta g_L^c$.
    Notice the vastly different scales in the three panels.
  }
\label{gLcn2HDM}
\end{figure}

\begin{figure}[ht]
  \begin{center}
    \includegraphics[width=1.0\textwidth]{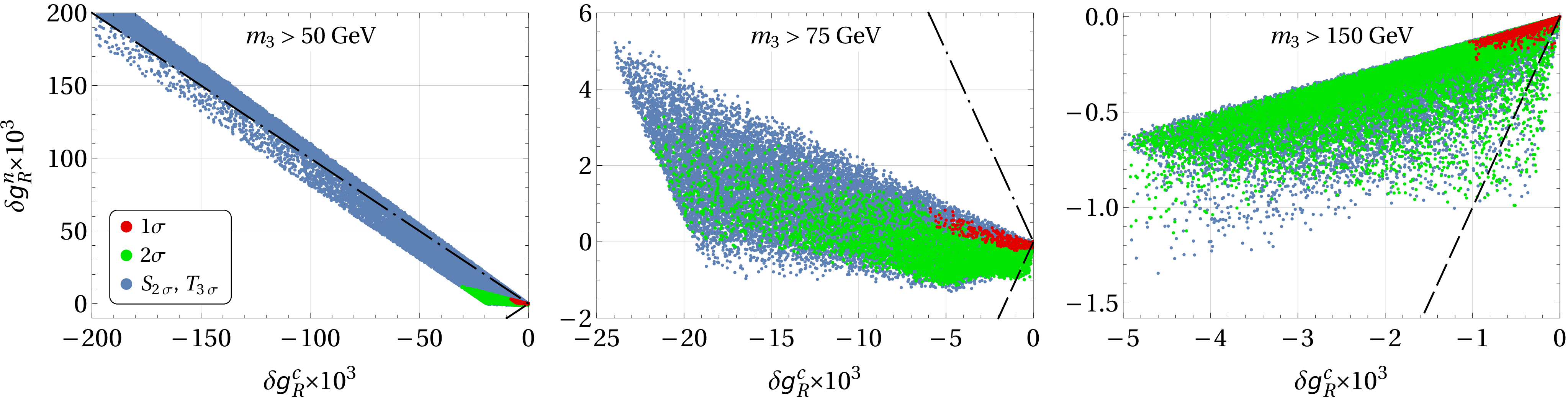}
  \end{center}
  \caption{Scatter plot of $\delta g_R^n$ \textit{versus}\/ $\delta g_R^c$
    in the aligned 2HDM.
    The displayed points and the colour code employed
    are the same as in figure~\ref{gLR2HDM}.
    The dashed straight lines mark the condition $\delta g_R^n = \delta g_R^c$
    and the dashed-dotted lines correspond to $\delta g_R^n = - \delta g_R^c$.
  }
\label{gRcn2HDM}
\end{figure}

\begin{figure}[ht]
  \begin{center}
    \includegraphics[width=1.0\textwidth]{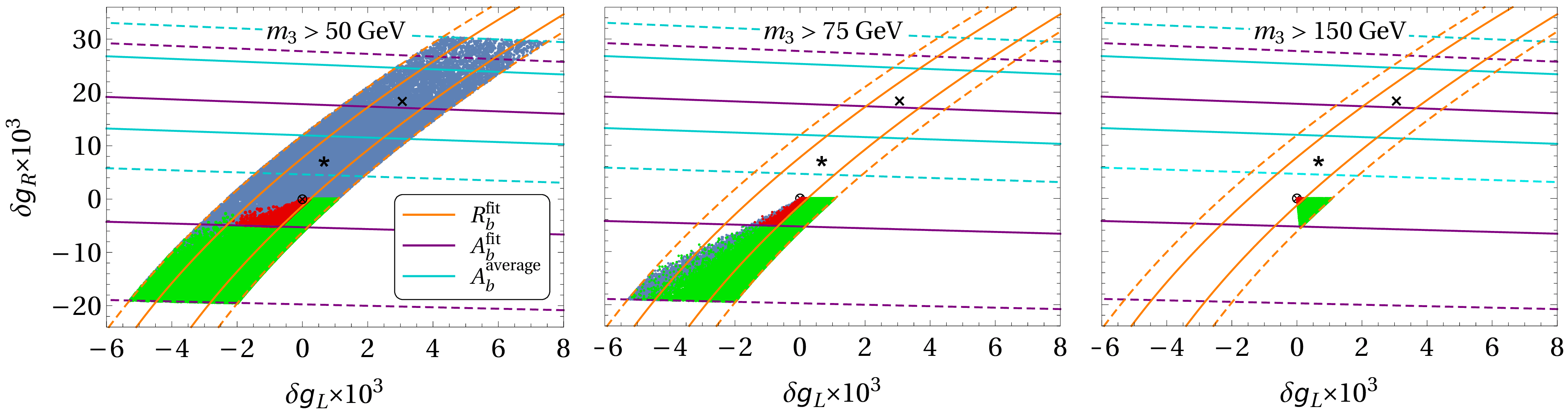}
  \end{center}
  \caption{Scatter plot of values of $\delta g_L$ and $\delta g_R$
    in the aligned 3HDM.
    All the conventions are the same as in figure~\ref{gLR2HDM}.
  }
\label{gLR3HDM}
\end{figure}

We depict in figure~\ref{gLR2HDM} the confrontation between experiment
and the values of $\delta g_L$ and $\delta g_R$ attainable in the aligned 2HDM.
One sees that,
if one forces the 2HDM to comply with the
$S$ and $T$-oblique parameter constraints~\eqref{ST},
then the 2HDM cannot achieve a better agreement with solution~1
for $g_L$ and $g_R$ than the SM;
in particular,
when one uses the $A_b$ value~\eqref{Abtrue},
the 2HDM cannot even reach the $2\sigma$ interval.
Only when one allows both for a laxer $S$- and $T$-oblique parameters constraints
and for a very low neutral-scalar mass $m_3 \lesssim 60$\,GeV
are the central values of both solutions 1$^\mathrm{fit}$
and 1$^\mathrm{average}$ attainable.
In the right panel of figure~\ref{gLR2HDM} one sees that,
if both new neutral scalars of the 2HDM have masses larger than 150\,GeV,
then the fit to solution~1 is never better than in the SM case,
even if one does not take into account
the $S$ and $T$-parameter constraints.\footnote{
  We want to emphasize that
  the constraint on the oblique parameter $S$ does not modify
  most of our figures much (notable exceptions are the blue areas
  in the left panels of figures~\ref{gLR2HDM}--\ref{gLR3HDM}); usually (but not always!),
  the points that comply with all other constraints
  also comply with the $S$ ones.
  The oblique parameter $S$ does not affect as much models with new scalars
  as models with new fermions,
  like for instance the ones in ref.~\cite{gori}.}

In figure~\ref{gLcn2HDM} we display the same points as in figure~\ref{gLR2HDM},
now distinguishing the neutral-scalar contribution to $\delta g_L$
from the charged-scalar contribution to the same quantity.
The same exercise is performed in figure~\ref{gRcn2HDM}
for the contributions to $\delta g_R$.
In the left panels of figures~\ref{gLcn2HDM} and~\ref{gRcn2HDM}
one can see that the agreement of some blue points with solution~1$^\mathrm{fit}$
is obtained not just by using very light neutral scalars
and laxer oblique parameters $S$ and $T$,
but also through a fine-tuning where
large neutral-scalar and charged-scalar contributions
almost cancel each other.
In the right panels of those figures one sees that,
when both neutral scalars have masses above 150\,GeV,
the signs of the neutral-scalar and charged-scalar contributions
are the same---this explains the agreement worse than in the SM
observed in figure~\ref{gLR2HDM}.

One also sees in figures~\ref{gLcn2HDM} and~\ref{gRcn2HDM} that
the neutral-scalar contributions $\delta g_L^n$ and $\delta g_R^n$
are often comparable in size to,
or even larger than,
the charged-scalar contributions $\delta g_L^c$ and $\delta g_R^c$,
respectively.
Thus,
the usual practice of taking into account just the charged-scalar contribution
may lead to erroneous results.

One might hope the situation of disagreement with experiment
to be milder in the 3HDM relative to the 2HDM,
but one sees in figure~\ref{gLR3HDM} that this hardly happens.
The agreement of the 3HDM with experiment
may be better than the one of the 2HDM, but
only in the case where very light neutral scalars exist.
We have checked that, just as in the 2HDM, the
better agreement occurs through an extensive finetuning
where $\delta g_L^n \approx - \delta g_L^c$
and $\delta g_R^n \approx - \delta g_R^c$.

\begin{figure}[ht]
  \begin{center}
    \includegraphics[width=1.0\textwidth]{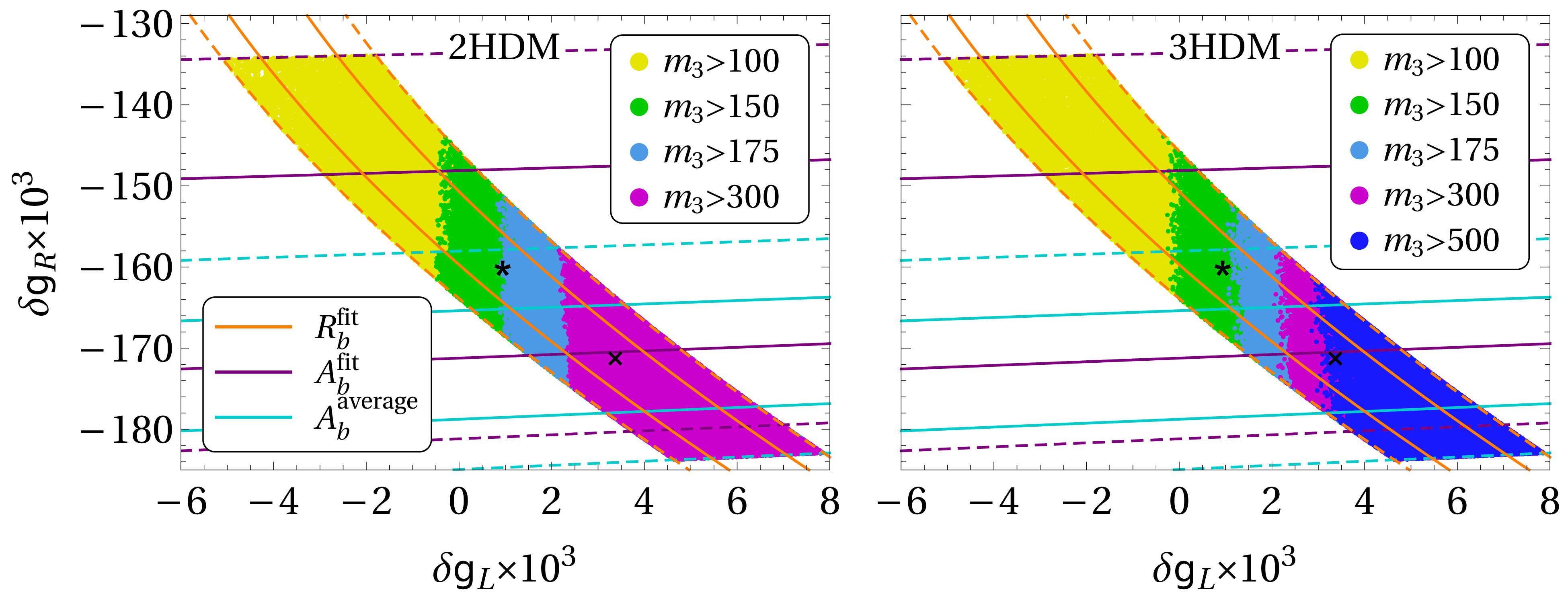}
  \end{center}
  \caption{
    Scatter plot of values of $\delta g_L$ and $\delta g_R$
    in the aligned 2HDM (left panel) and in the aligned 3HDM (right panel)   
    that suit the solution~2 for $R_b$ and $A_b$.
    All the points depicted comply
    with the $S$ and $T$-oblique parameters constraint of equation~\eqref{ST}. 
    A star marks the best-fit point of solution~2$^\mathrm{fit}$,
    and a cross the best-fit point of solution~2$^\mathrm{average}$.
    The orange lines mark the $1\sigma$ (full lines)
    and $2\sigma$ (dashed lines) boundaries of the region determined
    by the experimental value~\eqref{Rbfit};
    similarly,
    the violet lines correspond to the value~\eqref{Abfit}
    and the light-blue lines to the value~\eqref{Abtrue}.
    Blue points have new neutral scalars heavier than 500\,GeV,
    pink points have them heavier than 300\,GeV,
    light blue points have the lightest new neutral scalar    
    in between 175\,GeV and 300\,GeV,
    green points have it in between 150\,GeV and 175\,GeV,
    and yellow points have it between 100\,GeV and 150\,GeV.
  }
\label{print39}
\end{figure}

\begin{figure}[ht]
  \begin{center}
    \includegraphics[width=1.0\textwidth]{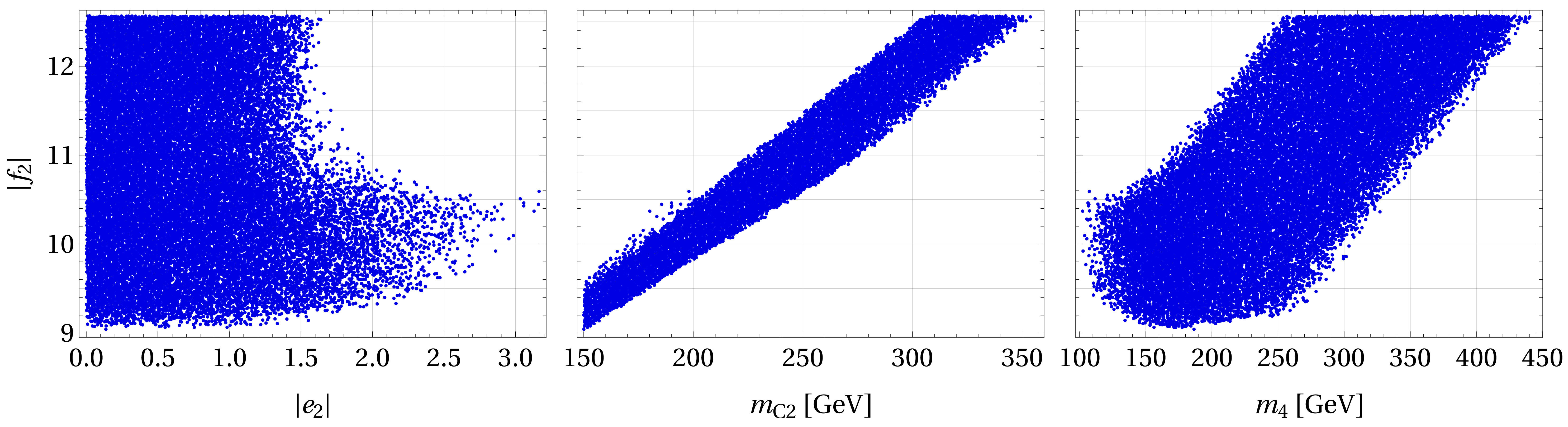}
  \end{center}
  \caption{Scatter plot of 2HDM points
    that obey the $S$- and $T$-oblique parameters constraints,
    fit solution~2$^\mathrm{average}$ at the 1$\sigma$ level,
    and have $m_3 > 100$\,GeV.
  }
\label{print39a}
\end{figure}

In figures~\ref{gLR2HDM}--\ref{gLR3HDM} we have tried,
and failed,
to make the fits of solution~1 in the 2HDM and in the 3HDM
better than in the SM.
Things are different with solution~2,
which the $n$HDM models can easily reproduce---with some caveats.
We remind the reader that in solution~2 the parameter $g_L$
is about the same as predicted by the SM,
but the parameter $g_R$ has sign opposite
to the one in the SM, \textit{viz.}\ $g_R \approx -0.08$ in solution~2
while $g_R \approx +0.08$ in the SM.
In the left panel of figure~\ref{print39} and in figure~\ref{print39a}
we see how the fit of solution~2
works out in the case of the 2HDM.
One sees that one can attain the 1$\sigma$ intervals and the best-fit points
both of solution 2$^\mathrm{fit}$ and of solution 2$^\mathrm{average}$,
but this requires
(1) the new scalars of the 2HDM to be lighter than 440\,GeV,
(2) the Yukawa coupling $f_2$ to be quite large,
and (3) the Yukawa coupling $e_2$ to be relatively small,
possibly even zero.
In practice,
the upper bound on the masses of the scalars
originates in the upper bound that unitarity imposes on $f_2$,
as seen in the middle panel of figure~\ref{print39a};
we have taken
(rather arbitrarily)
that upper bound to be
$\left| f_2 \right| < 4 \pi \approx 12.5$.
In the left panel of figure~\ref{print39a} one sees
that $\left| f_2 \right|$ must be larger than~9 anyway.
It is also clear from figure~\ref{print39} that,
the lighter the new scalars are allowed to be,
the easier it is to reproduce solution~2;
moreover,
it is easier to reproduce solution 2$^\mathrm{average}$,
\textit{viz.}\ with the value~\eqref{Abtrue} for $A_b$,
than solution 2$^\mathrm{fit}$,
\textit{viz.}\ with the value~\eqref{Abfit} for $A_b$,
because solution 2$^\mathrm{average}$ does not necessitate $m_3$ to be as low
as solution 2$^\mathrm{fit}$.

\begin{figure}[ht]
  \begin{center}
    \includegraphics[width=1.0\textwidth]{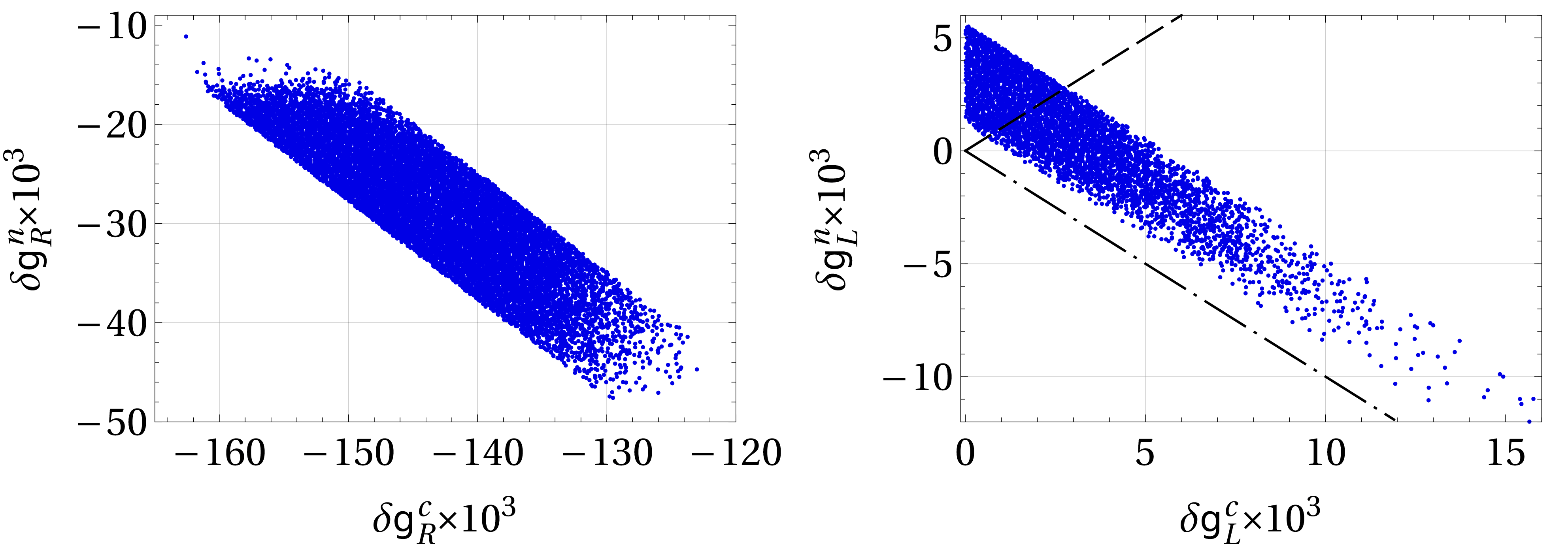}
  \end{center}
  \caption{
    Scatter plot of 3HDM points that obey the $S$- and $T$-oblique parameters constraints,
    fit solution~2$^\mathrm{average}$ at the 1$\sigma$ level,
    and have $m_3 > 100$\,GeV.
    Left panel: the neutral-scalar contribution
    to $\delta g_R$
    \textit{versus} the charged-scalar contribution to the same quantity.
    Right panel: the same for $\delta g_L$ instead of $\delta g_R$.
    In the right panel,
    the dashed line marks
    the condition $\delta g_L^n = \delta g_L^c$
    and the dashed-dotted line corresponds
    to $\delta g_L^n = - \delta g_L^c$.
  }
\label{print40}
\end{figure}

\begin{figure}[ht]
  \begin{center}
    \includegraphics[width=1.0\textwidth]{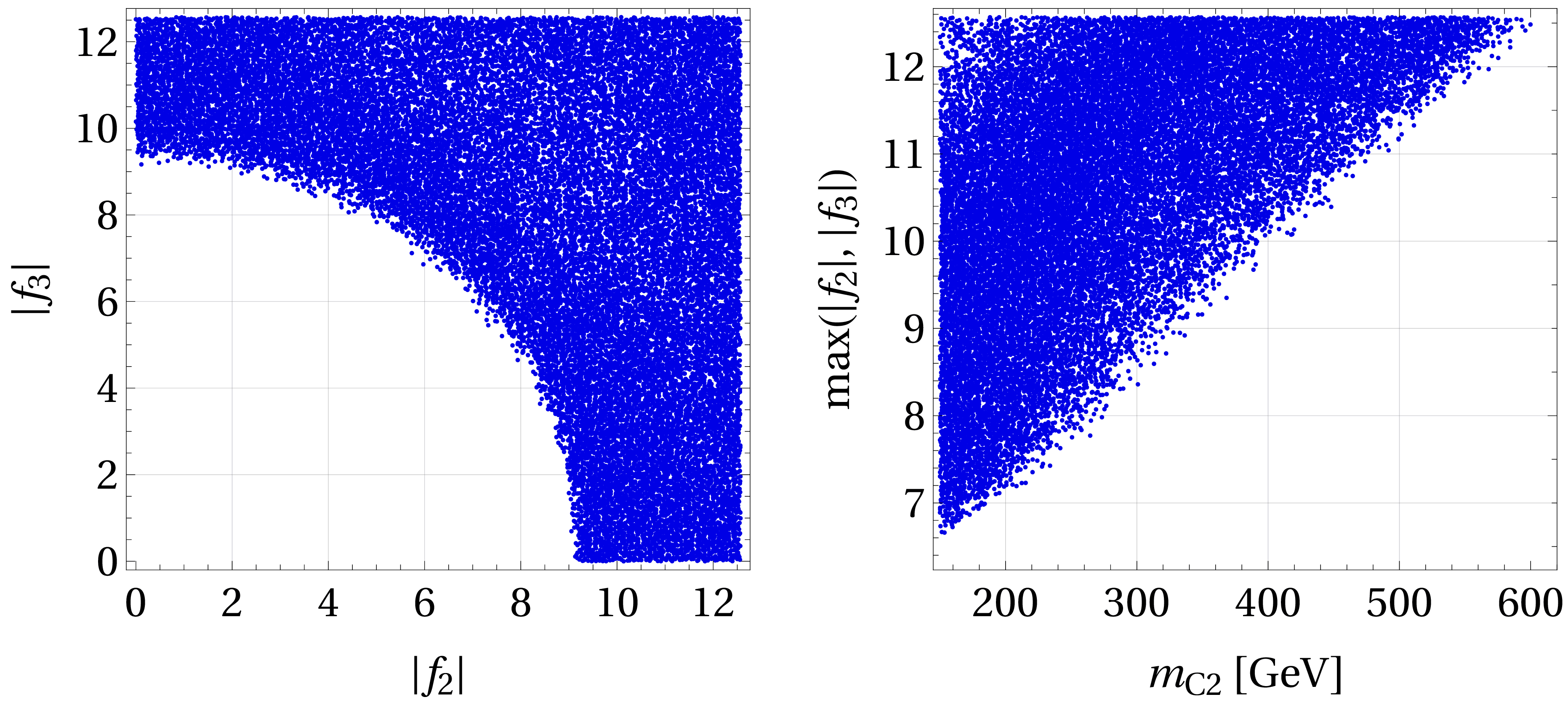}
  \end{center}
  \caption{
    Scatter plot of 3HDM points that obey the $S$- and $T$-oblique parameters constraints,
    fit solution~2$^\mathrm{average}$ at the 1$\sigma$ level,
    and have $m_3 > 100$\,GeV.
    Left panel: the modulus
    of the Yukawa coupling $f_3$ \textit{versus}\/
    the modulus of $f_2$.
    Right panel:
    the largest of the two Yukawa couplings $f_2$ and $f_3$
    \textit{versus} the mass of the lightest charged scalar.
  }
\label{print40a}
\end{figure}

In the right panel of figure~\ref{print39} and in figures~\ref{print40}
and~\ref{print40a}
we illustrate the fitting of solution~2 in the 3HDM.
Comparing the left and right panels of figure~\ref{print39},
we see that the 2HDM and the 3HDM give similar results,
but in the 3HDM it is possible to reach solution~2$^\mathrm{average}$
with larger masses of the new scalars.
Indeed,
in the 3HDM the lightest neutral scalar $m_3$ may be as heavy as 620\,GeV,
while in the 2HDM $m_3 < 420$\,GeV.
Like in figure~\ref{print39a},
in figures~\ref{print40} and~\ref{print40a}
we have used points that satisfy the $S$- and $T$-parameters
$1\sigma$ bounds~\eqref{ST},
that fall into the 1$\sigma$ intervals of $\delta g_L$ and $\delta g_R$
for solution 2$^\mathrm{average}$,\footnote{The fit of solution~2$^\mathrm{fit}$
  is not qualitatively different from the one of~2$^\mathrm{average}$;
  we concentrate on the latter just for the sake of simplicity.}
and that have $m_3 > 100$\,GeV.
In figure~\ref{print40} we display the charged- and neutral-scalar
contributions to $\delta g_L$ and $\delta g_R$.
One sees that solution~2 may be considered a finetuning,
with $\left| \delta g_R^c \right| \gg \left| \delta g_R^n,\,
\delta g_L^c,\, \delta g_L^n \right|$.
We stress once again that the neutral-scalar contributions
are as instrumental as the charged-scalar ones in obtaining decent fits.
In figure~\ref{print40a} we illustrate the moduli of the $f$ Yukawa couplings
and their relationship to the masses of the scalars.
One sees that there is a
bound $\sqrt{\left| f_2 \right|^2 + \left| f_3 \right|^2} \gtrsim 9$,
but each one of the Yukawa couplings  $f_2$ and $f_3$ may separately vanish.
One also observes that there is a simple straight-line correlation
between the maximum possible value for the mass of the lightest charged scalar,
$m_{C2}$,
and the minimum possible value for the largest of the Yukawa couplings
$f_2$ and $f_3$.
It is worth pointing out that
in the 3HDM, just as in the 2HDM, the masses $m_3$,
$m_4$,
and $m_{C2}$ must be low
(because of the unitarity upper bound on the $f$ Yukawa couplings),
but in the 3HDM the masses $m_5$,
$m_6$,
and $m_{C3}$ do not need to be low---they may be of order TeV.

\section{Conclusions} \label{sec:conclusions}

The Standard Model (SM) has a slight problem in fitting the $Z b \bar b$ vertex,
since it produces a $g_R$ smaller than
what is needed to reproduce the fit~\eqref{dez};
this discrepancy becomes larger
when one uses for $A_b$ the value~\eqref{Abtrue}.
In this paper we have found that this small problem can only
worsen when one extends the SM
through a $n$HDM.
This is because the contributions of the new scalars usually produce
a negative $\delta g_R$,
\textit{i.e.}\ they go in the wrong direction to alleviate the problem,
aggravating it instead.

There is one possible escape from this conclusion
if the extra neutral scalars of the $n$HDM are very light,
\textit{i.e.}\ lighter than the Fermi scale,
because the contribution of the neutral scalars to $\delta g_R$
may in that case be positive and partially compensate
for the inevitably negative contribution of the charged scalars.
This is a contrived effort,
though,
both because it is experimentally difficult
to accomodate very light neutral scalars and because,
from the theoretical side,
light neutral scalars together with heavy charged scalars
easily lead to a much-too-large
oblique parameter $T$.

In this paper we have
considered the possibility
that,
in $n$HDM models,
we might look instead at an alternative fit of the $Z b \bar b$ vertex,
wherein $g_R$ has the opposite sign from the one predicted by the SM.
This is what we have called ``solution~2'' in table~\ref{table_solutions}.
That solution necessitates a very large negative $\delta g_R$
(together with a small $\delta g_L$),
that may seem like a finetuning,
but is easy to obtain in a $n$HDM.
This solution,
though,
also works only if the new scalars are relatively light
and if at least one of the Yukawa couplings denoted $f_k$
in equation~\eqref{ef} is quite large,
\textit{viz.}\ larger than 9 or so.

\vspace*{5mm}

\paragraph{Acknowledgements:}
D.J.\ thanks the Lithuanian Academy of Sciences
for financial support through projects DaFi2019 and DaFi2021.
L.L.\ thanks
the Portuguese Foundation for Science and Technology
for support through CERN/FIS-PAR/0004/2019,
CERN/FIS-PAR/0008/2019,
PTDC/FIS-PAR/29436/2017,
UIDB/00777/2020,
and UIDP/00777/2020.

\newpage

\begin{appendix}

\setcounter{equation}{0}
\renewcommand{\theequation}{A\arabic{equation}}

\section{Definition and measurements of $R_b$ and $A_b$}
\label{appendixA}

The experimental quantities $R_b$ and $A_b$
are defined for $e^+ e^-$ collisions at the $Z^0$ peak,
\textit{i.e.}\ with $\sqrt{s} \approx m_Z$.
Let the quark $q$ ($q = u, d, s, c, b$) couple to the $Z^0$ as
\be
\mathcal{L}_{Zqq} = \frac{g}{c_w}\, Z_\mu\, \bar q \gamma^\mu
\left( g_{Lq} P_L + g_{Rq} P_R \right) q.
\ee
One has
\be
\label{72}
g_{Lq} = \frac{1}{2} - \frac{2 s_w^2}{3},
\quad \quad
g_{Rq} = - \frac{2 s_w^2}{3}
\ee
for $q = c$ and $q = u$,
and
\be
\label{73}
g_{Lq} = \frac{s_w^2}{3} - \frac{1}{2},
\quad \quad
g_{Rq} = \frac{s_w^2}{3}
\ee
for $q = b$,
$q = s$,
and $q = d$.
The probability that one produces a $q \bar q$ pair
in an $e^+ e^-$ collision at the $Z^0$ peak is,
in the absence of QCD,
QED,
and mass corrections proportional to
\be
\label{71}
s_q = 2 \left( g_{Lq}^2 + g_{Rq}^2 \right).
\ee
One finds from equations~\eqref{72}--\eqref{71} that
\be
\label{200}
s_c + s_u + s_s + s_d =
2 - 4 s_w^2 + \frac{40}{9}\, s_w^4 \approx 1.32827.
\ee
The experimental definition of $R_b$ is
\be
R_b = \frac{\Gamma_{b \bar b}}{\Gamma_\mathrm{hadrons}}
\ee
in $e^+ e^-$ collisions at the $Z^0$ peak;
thus,
$R_b$ is the fraction of the produced hadrons that contain a $b \bar b$ pair.
Clearly,
in the absence of QCD,
QED,
and mass corrections,
\bs
\ba
R_b &=& \frac{s_b}{s_b + s_c + s_u + s_s + s_d}
\label{75} \\
&=& \frac{9 - 12 s_w^2 + 8 s_w^4}{45 - 84 s_w^2 + 88 s_w^4}
\approx 0.21937.
\label{76}
\ea
\es
When one includes QCD,
QED,
and mass corrections
equation~\eqref{75} gets substituted by equation~\eqref{rvufido}
and $R_b$ decreases from the value in equation~\eqref{76} to 
the SM prediction~\cite{freitas} 0.21581.
Similarly,
$s_c + s_u + s_s + s_d$ becomes 1.3184
instead of 1.32827 as in equation~\eqref{200}.

If the mass of the bottom quark was zero,
equation~\eqref{v7fr895} would read
\be
A_b = \frac{g_L^2 - g_R^2}{g_L^2 + g_R^2}
= \frac{9 - 12 s_w^2}{9 - 12 s_w^2 + 8 s_w^4} \approx 0.94059
\ee
at the tree level.
The quantity $A_b$ was accessed at LEP~1 through the forward--backward asymmetry
of the produced $b \bar b$ quark pairs,
\be
\label{bvufiof}
A_\mathrm{FB}^{0,b} = \frac{3}{4}\, A_e A_b,
\ee
where $A_e$,
that is
\be
A_e = \frac{1 - 4 s_w^2}{1 - 4 s_w^2 + 8 s_w^4} \approx 0.21065
\ee
at the tree level,
can be extracted from other experiments.
Equation~\eqref{bvufiof} is the limit of
\be
A^{P_e,b}_\mathrm{FB} = \frac{3}{4}\, \frac{A_e + P_e}{1 + A_e P_e}\, A_b
\ee
when the polarization $P_e$ of the electron beam is zero.
The SLD Collaboration has used polarized beams ($P_e = 1$)
and therefore it could directly access
\be
\frac{3}{4}\, A_b = \frac{\sigma_{L\mathrm{F}} + \sigma_{R\mathrm{B}}
  - \sigma_{L\mathrm{B}} - \sigma_{R\mathrm{F}}}{\sigma_{L\mathrm{F}}
  + \sigma_{R\mathrm{B}} + \sigma_{L\mathrm{B}} + \sigma_{R\mathrm{F}}},
\ee
where the subscripts $L$ and $R$ refer to the electron's polarization
and the subscripts F and B refer to the forward or backward direction
of travel of the final-state bottom quarks.

The value of $A_b$ obtained from the SLD measurement
is $A_b^\mathrm{fit} = 0.923 \pm 0.020$ and is $0.6 \sigma$
below the SM value~\cite{RPP}. 
However,
this good agreement only applies to the overall fit of many observables.
Extracting $A_b$ from $A_{\mathrm{FB}}^{0,b}$ when $A_e = 0.1501 \pm 0.0016$
leads to $A_b = 0.885 \pm 0.0017$,
which is $2.9 \sigma$ below the SM prediction.
The combined value $A_b = 0.901 \pm 0.013$
deviates from the SM value by $2.6 \sigma$.
These discrepancies in $A_b$ could be an evidence of New Physics,
but they could also be due to a statistical fluctuation
or another experimental effect in one of asymmetries;
more precise experiments are needed. 

A direct measurement of the $Z b \bar b$ couplings at the LHC
is challenging because of the large backgrounds
in the detection of a $Z^0$ decaying into a bottom quark--antiquark pair.
A recent study~\cite{yan} has proposed a novel method
to probe the anomalous $Z b \bar b$ couplings
through the measurement of the cross section
of the associated production $gg \to Zh$ at the High Luminosity LHC. 

Lepton colliders of the next generation,
{\it vg.}\ the CEPC,
ILC,
or FCC-ee offer great opportunities for further studies
of the $Z b \bar b$ vertex,
because they could collect a large amount of data around the $Z^0$ pole.
In the analysis~\cite{Gori:2015nqa}
there is a list of the observables that are most important
for improving the constraints on the $Z b \bar b$ coupling,
and of the expected precision reach
of those three proposed future $e^+e^-$ colliders.
These estimates,
for the observables directly related to the $Z b \bar b$ coupling, 
are summarized in table~\ref{precisions}.
\begin{table}[t]\small
  \centering
  \begin{tabular}{|c|c||c|c|c|c|} \hline
    \multirow{2}{*}{Observable} & Current &
    \multicolumn{4}{c|}{Precision}
    \\ \cline{3-6}
    & measurement & Current & CEPC & ILC & FCC-ee
    \\ \hline\hline
    \multirow{2}{*}{$R_b$} &
    \multirow{2}{*}{0.21629} &
    0.00066  & 0.00017 & 0.00014 & 0.00006
    \\
    & & (0.00050) & (0.00016) & & (0.00006)
    \\ \hline
    \multirow{2}{*}{$A_{\mathrm{FB}}^{0,b}$} &
    \multirow{2}{*}{0.0996} &
    0.0016 & 0.00015 & &
    \\
    & & (0.0007) & (0.00014) & &
    \\ \hline
    \multirow{2}{*}{$A_b$} &
    \multirow{2}{*}{0.923} &
    0.020 & & 0.001 & 0.00021
    \\
    & & & & & (0.00015)
%    \\ \hline
%    %
%    \multirow{2}{*}{$A^0_{LR}$} &
%    \multirow{2}{*}{0.15138} &
%    0.00216 & & 0.0001 & 0.000021
%    \\
%    & & (0.0011) & & (0.0001) & (0.000015)
    \\ \hline \hline
    \# of $Z^0$s &
    $\sim 2 \times 10^7$ &
    \phantom{$\frac{1^0}{1^0}$} &
    $\sim 2 \times 10^9$ &
    $\sim 10^9$ &
    $\sim 10^{12}$
    \\ \hline
  \end{tabular}
  % \vspace*{-5mm}
  \caption{The estimated precision reach
    for $Z b \bar b$ observables at future $e^+ e^-$ colliders
    according to ref.~\cite{Gori:2015nqa}.
    The present result for each observable is shown in the second column.
    The third column shows the $\sigma$ of the present measurements
    at LEP and SLC,
    respectively,
    while the other columns show the estimates
    of the precision reach for the future colliders.
    In each entry,
    the number in the top line shows the total uncertainty
    and the number (in parenthesis) in the bottom line
    shows the systematic uncertainty.
    The last row shows the expected number of $Z^0$ events
    that will be collected.
    \label{precisions}}
\end{table}
We see that,
with an increase of precision of more than one order of magnitude,
a future collider has the potential
to solve the $A_{\mathrm{FB}}^{0,b}$ discrepancy found at LEP.  
If its results are SM-like,
a future lepton collider can provide strong constraints
on models beyond the SM;
if the $A_{\mathrm{FB}}^{0,b}$ discrepancy found at LEP
does come from New Physics,
then any of the three next-generation $e^+e^-$ colliders
will be able to rule out the SM
with more than $5\sigma$ significance~\cite{Gori:2015nqa}.

%%%%%%%%%%%%%%%%%%%%%%%%%%%%%%%%%%%%%%%%%%%%%%%%%%%%%%%%%%%%%%%%%%%%%%%%

%%%%%%%%%%%%%%%%%%%%%%%%%%%%%%%%%%%%%%%%%%%%%%%%%%%%%%%%%%%%%%%%%%%%%%%%

\setcounter{equation}{0}
\renewcommand{\theequation}{B\arabic{equation}}

\section{General formula for the neutral-scalar contribution}
\label{appendixB}

According to ref.~\cite{fontes},
the contributions to $\delta g_L$ and $\delta g_R$
of loops with internal lines of neutral scalars and bottom quarks
are the sums of three types of Feynman diagrams.
Thus,
\be
\label{sum}
\delta g_L^n = \delta g_L^n (a) + \delta g_L^n (b) + \delta g_L^n (c),
\quad
\delta g_R^n = \delta g_R^n (a) + \delta g_R^n (b) + \delta g_R^n (c).
\ee
Equations~(24),
(42),
and~(46) of ref.~\cite{fontes} inform us that
\bs
\label{cvuf859}
\ba
\delta g_L^n (a) &=& \frac{-i}{32 \pi^2} \sum_{l, l^\prime = 1}^{2n}
\mathcal{A}_{l l^\prime} \left( \mathcal{V}^T F \right)_l
\left( \mathcal{V}^\dagger F^\ast \right)_{l^\prime}\,
C_{00} \left( 0,\, m_Z^2,\, 0,\, 0,\, m_{l^\prime}^2,\, m_l^2 \right),
\\
\delta g_R^n (a) &=& \frac{-i}{32 \pi^2} \sum_{l, l^\prime = 1}^{2n}
\mathcal{A}_{l l^\prime} \left( \mathcal{V}^\dagger F^\ast \right)_l
\left( \mathcal{V}^T F \right)_{l^\prime}\,
C_{00} \left( 0,\, m_Z^2,\, 0,\, 0,\, m_{l^\prime}^2,\, m_l^2 \right),
\ea
\es
where $\mathcal{A}$ is the matrix defined in equation~\eqref{A},
and
\be
F = \left( \begin{array}{c} \left. \sqrt{2} m_b \right/ \! v \\*[0.5mm]
  f_2 \\ \vdots \\
  f_n \end{array} \right)
\ee
is a vector formed by Yukawa coupling constants.
Now,
\be
C_{00} \left( 0,\, m_Z^2,\, 0,\, 0,\, m_{l^\prime}^2,\, m_l^2 \right)
= C_{00} \left( 0,\, m_Z^2,\, 0,\, 0,\, m_l^2,\, m_{l^\prime}^2 \right),
\ee
while $\mathcal{A}_{l l^\prime} = - \mathcal{A}_{l^\prime l}$.
Therefore,
equations~\eqref{cvuf859} may be rewritten
\bs
\ba
\delta g_L^n (a) &=& \frac{1}{16 \pi^2} \sum_{l = 2}^{2n-1} \sum_{l^\prime = l+1}^{2n}
\mathcal{A}_{l l^\prime}\, \mathrm{Im} \left[ \left( \mathcal{V}^T F \right)_l
\left( \mathcal{V}^\dagger F^\ast \right)_{l^\prime} \right]
C_{00} \left( 0,\, m_Z^2,\, 0,\, 0,\, m_{l^\prime}^2,\, m_l^2 \right),
\hspace*{3mm}
\\
\delta g_R^n (a) &=& - \delta g_L^n (a),
\ea
\es
where we have dropped from the sum over the scalars
the Standard-Model contribution proportional to $\mathcal{A}_{12}$.

According to equations~(2),
(25),
and~(26) of ref.~\cite{fontes},
\bs
\label{bc}
\ba
\delta g_L^n (b) + \delta g_L^n (c) &=& \frac{1}{16 \pi^2}
\sum_{l=2}^{2n} \left| \left( \mathcal{V}^T F \right)_l \right|^2
\theta \left( m_l^2 \right),
\\
\delta g_R^n (b) + \delta g_R^n (c) &=& \frac{1}{16 \pi^2}
\sum_{l=2}^{2n} \left| \left( \mathcal{V}^T F \right)_l \right|^2
\lambda \left( m_l^2 \right),
\ea
\es
where
\bs
\ba
\theta \left( m_l^2 \right) &:=&
\frac{s_w^2}{6} \left[
  2\, C_{00} \left( 0,\, m_Z^2,\, 0,\, m_l^2,\, 0,\, 0 \right)
  - \frac{1}{2}
  \right. \no & & \left.
  - m_Z^2\, C_{12} \left( 0,\, m_Z^2,\, 0,\, m_l^2,\, 0,\, 0 \right)
  \vphantom{\frac{1}{2}} \right]
+ \left( \frac{s_w^2}{6} - \frac{1}{4} \right)
B_1 \left( 0,\, 0,\, m_l^2 \right),
\\
\lambda \left( m_l^2 \right) &:=&
\left( \frac{s_w^2}{6} - \frac{1}{4} \right) \left[
  2\, C_{00} \left( 0,\, m_Z^2,\, 0,\, m_l^2,\, 0,\, 0 \right)
  - \frac{1}{2}
  \right. \no & & \left.
  - m_Z^2\, C_{12} \left( 0,\, m_Z^2,\, 0,\, m_l^2,\, 0,\, 0 \right)
  \vphantom{\frac{1}{2}} \right]
+ \frac{s_w^2}{6}\, B_1 \left( 0,\, 0,\, m_l^2 \right).
\ea
\es

We now write
\be
\label{VVV}
\mathcal{V} = \mathcal{R} + i \mathcal{I},
\ee
where the $n \times 2n$ matrices $\mathcal{R}$ and $\mathcal{I}$
are real and satisfy
\be
\mathcal{R} \mathcal{R}^T = \mathcal{I} \mathcal{I}^T
= \mathbbm{1}_{n_d \times n_d}, \quad
\mathcal{R} \mathcal{I}^T = \mathcal{I} \mathcal{R}^T
= 0_{n_d \times n_d},
\ee
\textit{cf.}\ equation~\eqref{tildeV}.
From equation~\eqref{VVV},
\be
\label{imVV}
\mathcal{A} := \mathrm{Im} \left( \mathcal{V}^\dagger \mathcal{V} \right)
= \mathcal{R}^T \mathcal{I} - \mathcal{I}^T \mathcal{R}.
\ee
It follows that
\be
\sum_{l^\prime = 1}^{2n} \mathcal{A}_{l l^\prime}
\left( \mathcal{V}^\dagger F^\ast \right)_{l^\prime}
= - i \left( \mathcal{V}^\dagger F^\ast \right)_l,
\quad
\sum_{l = 1}^{2n} \left( \mathcal{V}^T F \right)_l \mathcal{A}_{l l^\prime}
= - i \left( \mathcal{V}^T F \right)_{l^\prime}.
\ee
Therefore,
from equations~\eqref{bc},
\bs
\ba
\delta g_L^n (b) + \delta g_L^n (c)
&=& \frac{1}{32 \pi^2} \left[
  \sum_{l=2}^{2n} \left( \mathcal{V}^T F \right)_l
  \left( \mathcal{V}^\dagger F^\ast \right)_l
  \theta \left( m_l^2 \right)
  \right. \no & & \left.
  + \sum_{l^\prime =2}^{2n} \left( \mathcal{V}^T F \right)_{l^\prime}
  \left( \mathcal{V}^\dagger F^\ast \right)_{l^\prime}
  \theta \left( m_{l^\prime}^2 \right) \right]
\\ &=& \frac{1}{32 \pi^2} \left\{
\sum_{l=2}^{2n} \left( \mathcal{V}^T F \right)_l
\left[ i \sum_{l^\prime = 1}^{2n} \mathcal{A}_{l l^\prime}
  \left( \mathcal{V}^\dagger F^\ast \right)_{l^\prime} \right]
\theta \left( m_l^2 \right)
\right. \no & & \left.
+ \sum_{l^\prime = 2}^{2n}
\left[ i \sum_{l = 1}^{2n} \left( \mathcal{V}^T F \right)_{l}
  \mathcal{A}_{l l^\prime} \right]
\left( \mathcal{V}^\dagger F^\ast \right)_{l^\prime}
\theta \left( m_{l^\prime}^2 \right) \right\}
\\ &=& \frac{i}{32 \pi^2} \sum_{l, l^\prime = 1}^{2n}
  \left( \mathcal{V}^T F \right)_l \mathcal{A}_{l l^\prime}
  \left( \mathcal{V}^\dagger F^\ast \right)_{l^\prime}
  \left[ \theta \left( m_l^2 \right)
  + \theta \left( m_{l^\prime}^2 \right) \right],
\\
\delta g_R^n (b) + \delta g_R^n (c)
&=& \frac{i}{32 \pi^2} \sum_{l, l^\prime = 1}^{2n}
\left( \mathcal{V}^T F \right)_l \mathcal{A}_{l l^\prime}
\left( \mathcal{V}^\dagger F^\ast \right)_{l^\prime}
\left[ \lambda \left( m_l^2 \right)
  + \lambda \left( m_{l^\prime}^2 \right) \right].
\ea
\es
Thus,
from equations~\eqref{sum} and~\eqref{cvuf859},
\bs
\label{68}
\ba
\delta g_L^n &=& \frac{-i}{32 \pi^2} \sum_{l, l^\prime = 1}^{2n}
\mathcal{A}_{l l^\prime} \left( \mathcal{V}^T F \right)_l
\left( \mathcal{V}^\dagger F^\ast \right)_{l^\prime}
\left[ C_{00} \left( 0,\, m_Z^2,\, 0,\, 0,\, m_{l^\prime}^2,\, m_l^2 \right)
  \right. \no & & \left.
  - \theta \left( m_l^2 \right)
  - \theta \left( m_{l^\prime}^2 \right) \right],
\\
\delta g_R^n &=& \frac{-i}{32 \pi^2} \sum_{l, l^\prime = 1}^{2n}
\mathcal{A}_{l l^\prime} \left( \mathcal{V}^T F \right)_l
\left( \mathcal{V}^\dagger F^\ast \right)_{l^\prime}
\left[ -C_{00} \left( 0,\, m_Z^2,\, 0,\, 0,\, m_{l^\prime}^2,\, m_l^2 \right)
  \right. \no & & \left.
  - \lambda \left( m_l^2 \right)
  - \lambda \left( m_{l^\prime}^2 \right) \right].
\ea
\es
We define the functions
\bs
\ba
h_L \left( m_{l^\prime}^2,\, m_l^2 \right)
&:=& - C_{00} \left( 0,\, m_Z^2,\, 0,\, 0,\, m_{l^\prime}^2,\, m_l^2 \right)
+ \theta \left( m_{l^\prime}^2 \right) + \theta \left( m_l^2 \right),
\\
h_R \left( m_{l^\prime}^2,\, m_l^2 \right)
&:=& C_{00} \left( 0,\, m_Z^2,\, 0,\, 0,\, m_{l^\prime}^2,\, m_l^2 \right)
+ \lambda \left( m_{l^\prime}^2 \right) + \lambda \left( m_l^2 \right).
\ea
\es
These functions are symmetric under the interchange of their two arguments:
\be
h_L \left( m_{l^\prime}^2,\, m_l^2 \right)
= h_L \left( m_l^2,\, m_{l^\prime}^2 \right),
\quad
h_R \left( m_{l^\prime}^2,\, m_l^2 \right)
= h_R \left( m_l^2,\, m_{l^\prime}^2 \right).
\ee
Moreover,
by utilizing equations~(21) of ref.~\cite{fontes} it is easy to show that,
although the functions  $\theta \left( m_l^2 \right)$,
$\lambda \left( m_l^2 \right)$,
and $C_{00} \left( 0,\, m_Z^2,\, 0,\, 0,\, m_{l^\prime}^2,\, m_l^2 \right)$
contain divergences,
the functions $h_L \left( m_{l^\prime}^2,\, m_l^2 \right)$
and $h_R \left( m_{l^\prime}^2,\, m_l^2 \right)$ do not.
From equation~\eqref{68} we obtain
\bs
\label{n}
\ba
\delta g_L^n &=& \frac{1}{16 \pi^2}
\sum_{l=2}^{2n - 1} \sum_{l^\prime = l+1}^{2n}
\mathcal{A}_{l l^\prime}\,
\mathrm{Im} \left[ \left( \mathcal{V}^\dagger F^\ast \right)_l
  \left( \mathcal{V}^T F \right)_{l^\prime} \right]
h_L \left( m_{l^\prime}^2,\, m_l^2 \right),
\\
\delta g_R^n &=& \delta g_L^n \left[ h_L \left( m_{l^\prime}^2,\, m_l^2 \right)
  \to h_R \left( m_{l^\prime}^2,\, m_l^2 \right) \right].
\ea
\es
Thus,
the functions $h_L$ and $h_R$ are crucial in the computation
of the neutral-scalar
contributions
to $\delta g_L$ and $\delta g_R$,
respectively.
Those functions were not explicitly defined in ref.~\cite{fontes},
even though they were utilized in that paper.

\end{appendix}

\end{document}